\documentclass[12pt]{article}
\usepackage{amsmath}
\usepackage{graphicx,psfrag,epsf}
\usepackage{enumerate}
\usepackage{natbib}

\usepackage{amsthm}
\usepackage{amsmath}
\usepackage{natbib}
\usepackage{enumitem}
\usepackage[colorlinks,citecolor=blue,urlcolor=blue,filecolor=blue,linkcolor=blue,backref=page]{hyperref}
\usepackage{graphicx}

\usepackage{bm}
\usepackage{amssymb}
\usepackage{algpseudocode}
\usepackage{listings}
\usepackage{array,multirow}
\usepackage{float}
\usepackage{booktabs}

\graphicspath{{img/}}

\DeclareMathOperator{\rank}{rank}

\newcommand{\plaintextref}[1]{#1}

\newtheorem{theorem}{Theorem}
\newtheorem{definition}{Definition}

\begin{document}

\def\spacingset#1{\renewcommand{\baselinestretch}%
{#1}\small\normalsize} \spacingset{1}


{
  \title{\bf Restricted Regression in Networks}
  \author{Ian Taylor\\
    Department of Statistics, Colorado State University\\
    and \\
    Kayleigh P. Keller \\
    Department of Statistics, Colorado State University\\
    and \\
    Bailey K. Fosdick\thanks{
    bailey.fosdick@cuanschutz.edu}\hspace{.2cm}\\
    Department of Biostatistics and Informatics, Colorado School of Public Health
    }
  \maketitle
}

\bigskip
\begin{abstract}
Network regression with additive node-level random effects can be problematic when the primary interest is estimating unconditional regression coefficients and some covariates are exactly or nearly in the vector space of node-level effects. We introduce the Restricted Network Regression model, that removes the collinearity between fixed and random effects in network regression by orthogonalizing the random effects against the covariates. We discuss the change in the interpretation of the regression coefficients in Restricted Network Regression and analytically characterize the effect of Restricted Network Regression on the regression coefficients for continuous response data. We show through simulation with continuous and binary response data that Restricted Network Regression mitigates, but does not alleviate, network confounding, by providing improved estimation of the regression coefficients. We apply the Restricted Network Regression model in an analysis of 2015 Eurovision Song Contest voting data and show how the choice of regression model affects inference.
\end{abstract}

\noindent%
{\it Keywords:}  network regression, confounding, restricted spatial regression, random effects

\newpage
\spacingset{1.45}

\section{Introduction}

Network data are measurements about the relationships between pairs of entities.
These network measurements can be visualized as measurements on the edges between nodes of a graph \citep{becker1995visualizing}. Examples of network data include relationships between potential borrowers on peer-to-peer lending platforms \citep{lee2022evaluating} or annual migration between countries \citep{aleskerov2017network}. Data are typically represented as a matrix, $\bm{Y}$, where $y_{ij}$ is the value in row $i$ and column $j$, for $i,j=1,\dots,n$. The value $y_{ij}$ is the measurement about the dyadic relationship between the sending node $i$ and the receiving node $j$. If $y_{ij} = y_{ji}$ for all $i,j$, then the network is called undirected, otherwise it is called directed.

Network regression uses covariates measured on the node pairs to model the dyadic relationships. An example of such a model is
\begin{align}
    y_{ij} &= g(z_{ij}) \label{eqn:intro-plain-model}, \\
    z_{ij} &= \bm{x}_{ij}^\top \bm{\beta} + \gamma_{ij}, \label{eqn:intro-plain-model-regression}
\end{align}
where $y_{ij}$ is the observed network measure, $g(\cdot)$ is a function mapping latent continuous values, $z_{ij}$, to the observed $y_{ij}$, $\bm{x}_{ij}$ is a $p$-vector of covariates related to node $i$, node $j$ or the relationship from node $i$ to node $j$, $\bm{\beta}$ is a $p$-vector of regression coefficients, and $\gamma_{ij}$ is random error. For a binary observation $y_{ij} \in \{0,1\}$ indicating the presence or absence of an edge from node $i$ to node $j$, a probit version of (\ref{eqn:intro-plain-model}) is given by setting
\begin{align}
    g(z_{ij}) &= I(z_{ij} > 0), \\
    \gamma_{ij} &\overset{\text{i.i.d}}{\sim} N(0, \sigma^2) \label{eqn:network-model-iid-errors},
\end{align}
where 
the function $I(\cdot)$ is the indicator function \citep{albert_chib_1993}. For the rest of this paper, we will use $z_{ij}$ to refer to a continuous latent variable resulting from the regression equation, and $y_{ij}$ to refer to observations related to $z_{ij}$ through a function, $g(\cdot)$, possibly the identity function. 

Network regression has found applications in medical meta-analysis \citep[e.g.,][]{li2018bayesian, gwon2020network}, analysis of international politics \citep[e.g.,][]{campbell2019latent}, and social networks \citep[e.g.,][]{CANTNER2006innovators}. Other methods for modeling network data include stochastic block models \citep{HOLLAND1983stochastic} and exponential-family random graph models \citep{ROBINS2007introduction}, which allow inference on latent aspects of the network structure such as clusters and density. While many different properties of a network may be of interest, our primary interest is inference for the regression coefficients, $\bm{\beta}$.

The network structure of the data creates dependence among the measurements. Observations $\{y_{ij}\}$ with one or more nodes in common, for example $y_{ij}$ and $y_{ij'}$, can be dependent due to their common node, $i$. If not all of this dependence is captured by the covariates, latent random effects can be used to account for excess variation \citep{HOLLAND1983stochastic, wang1987stochastic, hoffetal2002, li2002unified, hoff2005bilinear}. In this approach, the measurements are modeled as conditionally independent given the latent structure. Additive node random effects can be used to account for dependence in the observations,
\begin{align}
    \gamma_{ij} &= a_i + b_j + \varepsilon_{ij}, \label{eqn:additive-model} \\
    a_i &\sim g_a(\bm{\theta}_a), \quad b_j \sim g_b(\bm{\theta}_b), \\
    \varepsilon_{ij} &\overset{\text{i.i.d}}{\sim} N(0,\sigma^2).
\end{align}
The terms $a_i$ and $b_j$ are called sender and receiver random effects, respectively, and are meant to capture variation due to unobserved node factors, for example,  sociability and popularity in social networks. The random effects have distributions, $g_a$ and $g_b$, parameterized by parameters $\bm{\theta}_a$ and $\bm{\theta}_b$. These effects first appeared in the social relations model of \citet{warner1979}. 
Node random effects have also been incorporated in more complex network models such as popularity-adjusted block models \citep{sengupta2018block} and additive and multiplicative effects network models \citep{hoff2021additive}. An alternate approach is to model the correlation between measurements on dyads which share a node by the residual covariance matrix under an assumption of exchangeable errors \citep{marrs2022regression}.

A key difference between the model in (\ref{eqn:intro-plain-model-regression}) with the error structure in (\ref{eqn:network-model-iid-errors}) compared to the error structure in (\ref{eqn:additive-model}) is the interpretation of the regression parameters. With the error structure in (\ref{eqn:network-model-iid-errors}), the regression parameters are called the \textit{unconditional regression effects} because they capture the marginal effect of $\bm{X}$ on $\bm{z}$. With (\ref{eqn:additive-model}), 
the regression parameters are called the \textit{conditional regression effects} because their value is interpreted as the effect of $\bm{X}$ conditioned on the random effects $\bm{a}=(a_1,\dots,a_n)$ and $\bm{b}=(b_1,\dots,b_n)$. For the rest of this paper, we will distinguish between these interpretations by writing $\bm{\beta}$ for conditional regression effects and $\bm{\delta}$ for unconditional regression effects in (\ref{eqn:intro-plain-model-regression}).

In network-structured data, covariates can occupy the same linear space as the random effects $a_i$ or $b_j$. We define this collinearity as \textit{network confounding}. 
Despite potential impacts of bias due to network confounding, including random effects is typically desired to account for correlation between observations due to network dependence and to allow for more accurate uncertainty quantification for the regression effects. A method that mitigates network confounding would allow for accurate estimation of the unconditional regression effects, while using random effects to account for unobserved network-structured variability in the response. 

Confounding between covariates and random effects has been of significant interest in the spatial statistics literature, where it is referred to as \textit{spatial confounding} \citep{Clayton_Bernardinelli_Montomoli_1993, reich2006effects, Hodges2010}. A typical spatial model for continuous areal data is the Intrinsic Conditional Autoregressive \citep[ICAR;][]{besag1991bayesian} model:
\begin{align}
\bm{y} &= \bm{X}\bm{\beta} + \bm{\eta} + \bm{\varepsilon}, \label{eqn:general-spatial-unrestricted} \\
p(\bm{\eta}) &\propto \tau_s^{n-G}\exp\left(-0.5\tau_s\bm{\eta}^\top \bm{Q} \bm{\eta}\right), \\
\bm{\varepsilon} &\sim N(\bm{0}, \sigma^2\bm{I}).
\end{align}
The random effect $\bm{\eta}$ is intended to capture spatial variation in the response $\bm{z}$ not accounted for by the predictors $\bm{X}$. The random effect is regularized to have spatial variation via the matrix $\bm{Q}$, the Laplacian of the graph of neighboring areas. Spatial confounding occurs when a covariate is smoothly spatially varying, as this creates a scenario when there is collinearity or near collinearity between the fixed covariates $\bm{X}$ and spatial random effects $\bm{\eta}$ such that both the random effect and the covariate are attempting to capture similar structure. \citet{Hodges2010} noted that estimates of regression coefficients $\bm{\beta}$ can be dramatically affected by the inclusion or exclusion of random effects $\bm{\eta}$,  and introduced restricted spatial regression to resolve the confounding of the random effects and covariates by orthogonalizing $\bm{\eta}$ against $\bm{X}$. The restricted spatial regression equation is expressed as
\begin{equation}\bm{y} = \bm{X}\bm{\delta} + (\bm{I}-\bm{P}_X)\bm{\eta} + \bm{\varepsilon},\label{eqn:general-spatial-restricted}\end{equation}
where $\bm{P}_X = \bm{X}(\bm{X}^\top \bm{X})^{-1} \bm{X}^\top$ is the linear projection matrix onto the column space of $\bm{X}$. Restricted spatial regression has been studied extensively in the spatial statistics literature as a means to alleviate spatial confounding \citep[e.g.,][]{Hodges2010, hughes2013dimension, hanks2015restricted, khan2020restricted, zimmerman2022deconfounding}. \citet{hanks2015restricted} show that in a Bayesian setting, inference on both $\bm{\beta}$ and $\bm{\delta}$ can be achieved simultaneously by calculating $\bm{\delta}= \bm{\beta} + (\bm{X}^\top \bm{X})^{-1} \bm{X}^\top \bm{\eta}$ for each posterior sample. 

\begin{figure}
\begin{tabular}{ccc}
& Data & Dependence Graph \\
\rotatebox[origin=l]{90}{\parbox[t]{1.8in}{\centering Spatial Setting}} & 
\includegraphics[width=0.42\linewidth]{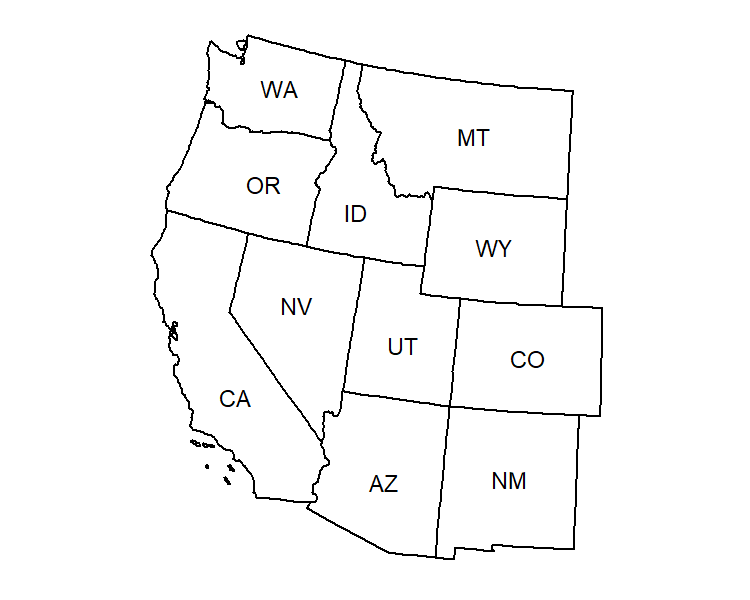} &
\includegraphics[width=0.42\linewidth]{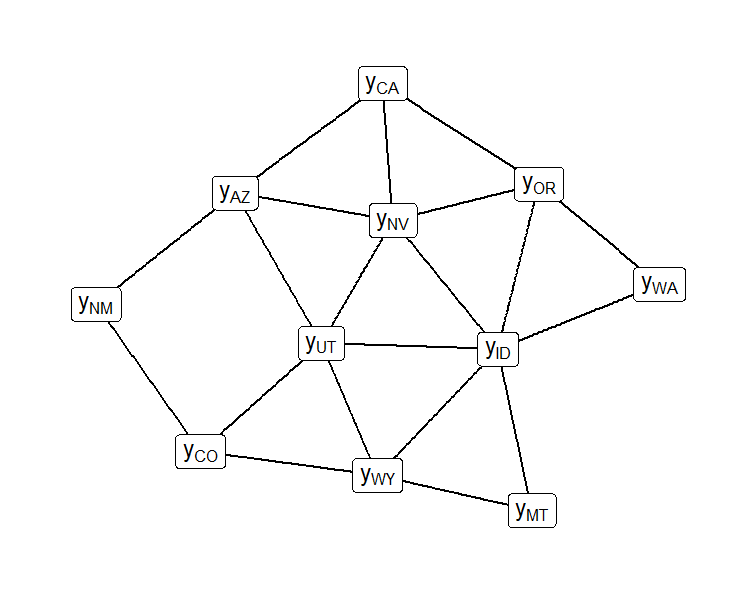} \\
& (A) Areal Data & (B) Spatial Dependence Graph \\
\rotatebox[origin=l]{90}{\parbox[t]{1.8in}{\centering Network Setting}} &
\includegraphics[width=0.42\linewidth]{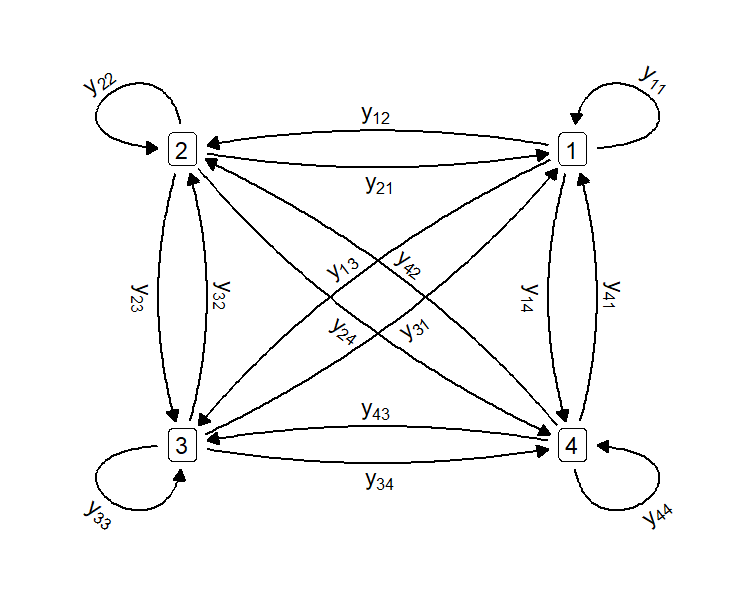} &
\includegraphics[width=0.42\linewidth]{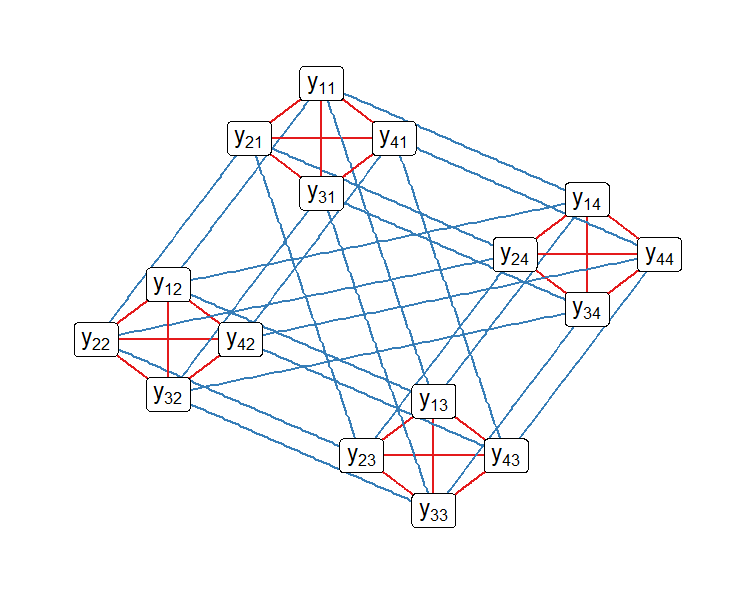} \\
& (C) Network Data & (D) Network Dependence Graph
\end{tabular}
\caption{Comparing spatial dependence to network dependence. (A): A map of 11 western states. (B): The  dependence graph corresponding to observations on each state in (A), where observations on states sharing a border are dependent. (C): A directed network with 4 nodes numbered 1 through 4, and observations measured on each directed edge. (D): The network dependence graph corresponding to observations in (C) . Each edge in the network is now a node in the dependence graph. There are two types of dependence - observations with a common sender (blue) and a common receiver (red). Both blue and red edges form distinct fully-connected clusters of observations.}
\label{fig:example-network-to-spatial}
\end{figure}

However, the dependence in network data is more complex than the dependence in spatial data. In the network setting, observations are made on the relationships between nodes, and shared nodes between two observations create dependence. This is in contrast to the areal spatial setting, where observations are made on discrete, disjoint areas and the neighbor relation between areas creates dependence (Figure \ref{fig:example-network-to-spatial}, A \& B). The additive random effects in (\ref{eqn:additive-model}), reflect two different ways in which network observations can be related: sharing a sender node (observations $y_{ij}$ and $y_{ij'}$) and sharing a receiver node (observations $y_{ij}$ and $y_{i'j}$). These two types of dependence result in a complex dependence structure (Figure \ref{fig:example-network-to-spatial}, C \& D). For each of these two types of dependence, the $n^2$ observations are divided into $n$ groups of $n$ observations which are all mutually dependent within a group. 

Spatial models in which regions can be neighboring in two distinct ways have been studied in \citet{reich2007spatial} for the case of periodontal health measurements. Measurements were considered to be neighboring other measurements on the same tooth, or measurements on an adjacent tooth, but these kinds of neighbor relations were considered to create different kinds of dependence between measurements. The dependence in the periodontal health measurements was modeled using the random effect prior
\begin{equation}p(\bm{\eta}) \propto c(\tau_1, \tau_2)\exp\{-0.5 \bm{\eta}^\top (\tau_1 \bm{Q}_1 + \tau_2 \bm{Q}_2) \bm{\eta}\},\label{eqn:spatial-two-neighbor-effect}\end{equation}
where $\bm{Q}_1$ and $\bm{Q}_2$ are Laplacians for each neighbor relation. However, the distribution in (\ref{eqn:spatial-two-neighbor-effect}) cannot represent the same dependence as the sender and receiver random effects in (\ref{eqn:additive-model}) because the distribution of $\bm{\eta}$ is full rank, while the distribution of the vectorized $(a_i+b_j)$ is not full rank.

In this paper we introduce Restricted Network Regression as a method that mitigates network confounding. We approach network regression in a Bayesian framework, estimating parameters using their posterior distributions given the data. We characterize the posterior mean and variance of regression parameters in Restricted Network Regression with continuous data and show through simulation that Restricted Network Regression mitigates network confounding for continuous and binary data. Specifically, we demonstrate that a Restricted Network Regression model results in smaller bias and posterior credible intervals that are more appropriately calibrated to capture the generative parameter values than corresponding network regression models without random effects or with non-restricted random effects.

The remainder of this paper is organized as follows. Section \ref{sec:network-confounding} defines network confounding and requirements for methods to ``alleviate'' and ``mitigate'' network confounding. Section \ref{sec:restricted-network-regression} introduces Restricted Network Regression, characterizes the collinearity of effects within the Restricted Network Regression model, and provides theorems about the posterior distribution of the regression parameters in the continuous Restricted Network Regression model. Section \ref{sec:simulation-study} describes the results of a simulation study involving both continuous and binary network regression. Section \ref{sec:eurovision-analysis} is a case study of Eurovision Song Contest voting data showing the changes in inference that occur when using a Restricted Network Regression approach, relative to models without random effects and non-restricted random effect models. Finally, Section \ref{sec:discussion} closes with a discussion.

\section{Network Confounding} \label{sec:network-confounding}

Network confounding is the collinearity between fixed effects and network-structured random effects in a network regression model. This collinearity creates difficulty in estimating regression parameters by introducing bias in estimates and increasing their posterior variance. In this section we explore network confounding in more detail and define conditions for a method to alleviate or mitigate network confounding. 

Consider the network regression model,
\begin{align}
    y_{ij} &= g^{-1}(z_{ij}), \ \ 1 \leq i,j \leq n, \\
    \bm{z} &= \bm{X}\bm{\beta} + \bm{A}\bm{a} + \bm{B}\bm{b} + \bm{\varepsilon} , \label{eqn:network-model-unrestricted-start}\\
    a_i &\overset{\text{i.i.d.}}{\sim} g_a(\bm{\theta_a}), \\
    b_j &\overset{\text{i.i.d.}}{\sim} g_b(\bm{\theta_b}), \\
    \bm{\varepsilon} &\sim N(\bm{0},\sigma^2_e \bm{I}), \label{eqn:network-model-unrestricted-end} 
\end{align}
where $\bm{z}$ is an $n^2$-vector of latent continuous responses, $\bm{X}$ is an $n^2 \times p$ fixed matrix of covariates, $\bm{\beta}$ is a $p$-vector of regression parameters, and $\bm{a}$ and $\bm{b}$ are $n$-vectors composed of sender and receiver random effects, respectively, for each node in the network. The matrices $\bm{A}$ and $\bm{B}$ are $n^2 \times n$ matrices of zeros and ones which broadcast the elements of $\bm{a}$ and $\bm{b}$ into the appropriate rows of $\bm{z}$ depending on the sender and receiver of each dyad. To match the vectorized form of $\bm{z}$, we write $\bm{y}$ as an $n^2$-vector of the observed network data.

Now consider partitioning $\bm{X} = [\bm{1}\ \bm{X}_s\ \bm{X}_r\ \bm{X}_d]$ into an intercept, an $n^2 \times p_s$ matrix of sender covariates $\bm{X}_s$, an $n^2 \times p_r$ matrix of receiver covariates $\bm{X}_r$, and an $n^2 \times p_d$ matrix of dyadic covariates $\bm{X}_d$ as done in \citet{hoff2021additive}, where $p_s$, $p_r$, and $p_d$ are the number of sender, receiver, and dyadic covariates, respectively, and $p = 1 + p_s + p_r + p_d$. Similarly, partition $\bm{\beta} = (\beta_0\ \bm{\beta}_s^\top\ \bm{\beta}_r^\top\ \bm{\beta}_d^\top)^\top$. The column space of the matrix $\bm{A}$ contains all $n^2$-vectors that have repeated values for dyads with a common sender. Similarly, the column space of the matrix $\bm{B}$ contains all $n^2$-vectors that have repeated values for dyads with a common receiver. Since the columns of $\bm{X}_s$ are sender covariates, every column of $\bm{X}_s$ is in the column space of $\bm{A}$. Therefore, we can write $\bm{X}_s = \bm{A}\bm{X}_s'$, where $\bm{X}_s'$ is an $n\times p_s$ matrix created by collapsing equivalent rows of $\bm{X}_s$. Since the columns of $\bm{X}_r$ are receiver covariates, every column of $\bm{X}_r$ is in the column space of $\bm{B}$, allowing us to write $\bm{X}_r = \bm{B}\bm{X}_r'$. The model in (\ref{eqn:network-model-unrestricted-start}) can be written
\begin{align}
    \bm{z} &= [\bm{1}\ \bm{X}_d](\beta_0\ \bm{\beta}_d^\top)^\top + \bm{X}_s \bm{\beta}_s + \bm{X}_r \bm{\beta}_r + \bm{A}\bm{a} + \bm{B}\bm{b} + \bm{\varepsilon}, \\
    &= [\bm{1}\ \bm{X}_d](\beta_0\ \bm{\beta}_d^\top)^\top + \bm{A}(\bm{X}_s'\bm{\beta}_s + \bm{a}) + \bm{B}(\bm{X}_r'\bm{\beta}_r + \bm{b}) + \bm{\varepsilon}.
\end{align}

From this, we can see that $\bm{\beta}_s$ and $\bm{a}$ are confounded in the sense that they occupy the same linear space in the response, i.e., ${\cal C}(\bm{X}_s) \subset {\cal C}(\bm{A})$, where ${\cal C}$ is the column space operator. Similarly, $\bm{\beta}_r$ and $\bm{b}$ are also confounded. 
By restricting the random effects to be orthogonal to the fixed effects, Restricted Network Regression (introduced formally in the next section) fixes this collinearity by removing the intersection of the column spaces of $\bm{X}$, $\bm{A}$, and $\bm{B}$.

We now distinguish between two types of methods: those that \textit{alleviate} network confounding and those that \textit{mitigate} network confounding. We adapt a definition from the spatial statistics literature to define what it means for a method to alleviate network confounding:

\begin{definition}
A network regression method modeling network data, $\bm{y}$, with unconditional regression parameters, $\bm{\delta}$, which results in posterior mean $\mathrm{E}[\bm{\delta}|\bm{y}]$ and marginal posterior variances $\mathrm{Var}(\delta_\ell|\bm{y}), \ell=1,\dots,p$ \underline{alleviates} network confounding if the following conditions are met:
\begin{enumerate}
    \item $\mathrm{E}[\bm{\delta}|\bm{y}] = \mathrm{E}[\bm{\delta}_{NN}|\bm{y}]$
    \item $\mathrm{Var}(\delta_{NN,\ell}|\bm{y}) \leq \mathrm{Var}(\delta_\ell|\bm{y}) \leq \mathrm{Var}(\beta_{Network,\ell}|\bm{y})$ for $\ell=1,\dots,p$
\end{enumerate}
Here $\delta_{NN,\ell}^X$ are the unconditional regression coefficients of the corresponding network model without network-structured random effects and $\beta_{Network,\ell}^X$ are the conditional regression coefficients from a model with non-restricted network-structured random effects.
\label{def:alleviates-network-confounding}
\end{definition}

Definition \ref{def:alleviates-network-confounding} is adapted from the definition of spatial confounding in \citet{khan2020restricted}.
This definition of alleviating network confounding reflects the intuition that models with network confounding will result in excess uncertainty on regression parameter estimates, reflected in large posterior variances of those parameters and that a method that alleviates network confounding should have lower posterior variances than one exhibiting network confounding. At the same time, a model that alleviates network confounding should model the unconditional effects of the fixed covariates, and so is expected to have the same posterior means as the model without network-structured random effects.

Alleviation of network confounding has useful interpretation in terms of parameter uncertainty, however, it does not relate directly to the accuracy of estimates made using models with network confounding. Also, Definition \ref{def:alleviates-network-confounding} provides no way to compare two models if neither alleviates network confounding. For these reasons, we introduce an alternative notion of network confounding mitigation. 
We expect a method that mitigates network confounding relative to another approach to produce better estimates and uncertainty quantification of the unconditional regression effects $\bm{\delta}$. We give a more precise definition of this mitigation:

\begin{definition}
    A network regression method modeling network data, $\bm{y}$, with unconditional regression parameters, $\bm{\delta}$, which results in posterior mean $\mathrm{E}[\bm{\delta}|\bm{y}]$, denoted $\bm{m}$, and marginal posterior $100c\%$ credible intervals $I_{c,\ell}$, $0 < c < 1$, for components $\delta_\ell$, \underline{mitigates} network confounding relative to another method which produces $\bm{m}^\prime$ and $I_{c,\ell}^\prime$, if for true unconditional regression effects $\bm{\delta}^\ast$,
    \begin{enumerate}
        \item $\left|\mathrm{E}\left[m_\ell - \delta_\ell^\ast\right]\right| \leq \left|\mathrm{E}\left[m^\prime_\ell - \delta_\ell^\ast\right]\right|$ for $\ell=1,\dots,p$,
        \item $\left|\mathrm{P}\left(\delta_\ell^\ast \in I_{c,\ell}\right) - c\right| \leq \left|\mathrm{P}\left(\delta_\ell^\ast \in I_{c,\ell}^\prime\right) - c\right|$  for $\ell=1,\dots,p$,
    \end{enumerate}
    where the expectation in item 1 and probability in item 2 are taken over $\bm{y}$.
    \label{def:mitigates-network-confounding}
\end{definition}

Definition \ref{def:mitigates-network-confounding} combines two useful model evaluations, bias and credible interval coverage, into a comparison which can be used to evaluate the relative improvement of one model over another in the presence of network confounding. Even if a model does not meet the requirements of Definition \ref{def:alleviates-network-confounding}, it can be compared to other models using Definition \ref{def:mitigates-network-confounding} to assess its mitigation of network confounding.

\section{Restricted Network Regression} \label{sec:restricted-network-regression}

For the network regression model given in  (\ref{eqn:network-model-unrestricted-start}) - (\ref{eqn:network-model-unrestricted-end}), we propose the following Restricted Network Regression model:
\begin{align}
\bm{z} &= \bm{X}\bm{\delta} + (\bm{I}-\bm{P}_X)(\bm{A}\bm{a} + \bm{B}\bm{b}) + \bm{\varepsilon}, \label{eqn:restricted-network-regression-model} \\
a_i &\overset{\text{i.i.d.}}{\sim} g_a(\bm{\theta_a}), \\
b_j &\overset{\text{i.i.d.}}{\sim} g_b(\bm{\theta_b}), \\
\bm{\varepsilon} &\sim \mathrm{N}(\bm{0},\sigma^2 \bm{I}_{n^2}), 
\end{align}
where $\bm{z}$, $\bm{X}$, $\bm{a}$, $\bm{b}$, $\bm{A}$, $\bm{B}$, and $\bm{P}_X$ are as described earlier. This model is distinguished from the model in (\ref{eqn:network-model-unrestricted-start})-(\ref{eqn:network-model-unrestricted-end}) by the application of the projection matrix $\bm{I}-\bm{P}_X$ projecting the random effects orthogonal to the column space of $\bm{X}$, and the interpretation of the regression effects, $\bm{\delta}$, as unconditional on the values of the random effects. With $\bm{\delta}$ related to the unconditional regression parameters, $\bm{\beta}$ by $\bm{\delta} = \bm{\beta} + (\bm{X}^\top \bm{X})^{-1}\bm{X}^\top(\bm{A}\bm{a} + \bm{B}\bm{b}),$ 
the regression equation (\ref{eqn:restricted-network-regression-model}) is equivalent to (\ref{eqn:network-model-unrestricted-start}).

\subsection{Properties of Continuous Restricted Network Regression Posterior Distributions} \label{sec:rsr-theory}

In this section, we provide relationships between the posterior means and variances of Restricted Network Regression and a model without random effects. With these relationships, we show that Restricted Network Regression with a continuous response does not satisfy the conditions of Definition \ref{def:alleviates-network-confounding}, and therefore does not alleviate network confounding. We also relate these theoretical results to recent results in the spatial statistics literature by \citet{khan2020restricted}, which showed a similar result in spatial regression.

To understand the behavior of the Restricted Network Regression model with respect to network confounding, we give an expression for the posterior mean and variance of $\bm{\delta}$ in the Restricted Network Regression model.

\begin{theorem}
In a continuous Restricted Network Regression model with two additive normally distributed node-level random effects,
\begin{align}
    \bm{y} &= \bm{X}\bm{\delta} + (\bm{I} - \bm{P}_X)\bm{A}\bm{a} + (\bm{I} - \bm{P}_X)\bm{B}\bm{b} + \bm{\varepsilon}, \\
    p(\bm{\delta}) &\propto 1, \\
    \bm{a} &\sim N(\bm{0}, \sigma^2_a \bm{I}),\ \bm{b} \sim N(\bm{0}, \sigma^2_b \bm{I}), \bm{\varepsilon} \sim N(\bm{0}, \sigma^2_\varepsilon \bm{I}),
\end{align}
the posterior distribution of $\bm{\delta}$ will have mean and variance
\begin{align}
    \mathrm{E}[\bm{\delta}|\bm{y}] &= (\bm{X}^\top \bm{X})^{-1} \bm{X}^\top \bm{y}, \label{eqn:network-two-restricted-effects-posterior-mean} \\
    \mathrm{Var}(\bm{\delta} | \bm{y}) &= (\bm{X}^\top \bm{X})^{-1} \mathrm{E}[\sigma^2_\varepsilon | \bm{y}].
\end{align}

\label{thm:network-2randomeffects-1}
\end{theorem}

The proof of this theorem is provided in Appendix \plaintextref{A} of the Supplementary Material \citep{articlesupplement}. The prior distribution of $\sigma^2_\varepsilon$ is not specified for this theorem, but will partially determine $\mathrm{E}[\sigma^2_\varepsilon | \bm{y}]$. The posterior mean in (\ref{eqn:network-two-restricted-effects-posterior-mean}) is equal to the posterior mean from a model without network-structured random effects, as required by Definition \ref{def:alleviates-network-confounding}. Therefore Restricted Network Regression alleviates network confounding if and only if the inequality on the posterior variances in Definition \ref{def:alleviates-network-confounding} is true. We show this property of the posterior variances using a more general model with two restricted random effects.

\begin{theorem}
Consider the restricted regression model with two random effects,
\begin{align}
  \bm{y} &= \bm{X}\bm{\delta} + \bm{W}_1 \bm{\eta}_1 + \bm{W}_2 \bm{\eta_2} + \bm{\epsilon}, \\
  p(\bm{\delta}) &\propto 1, \\
  p(\bm{\eta}_1 | \tau_1) &\sim \tau_1^{\rank(\bm{F}_1)/2}\exp\left\{-\frac{\tau_1}{2}\bm{\eta}_1^\top \bm{F}_1 \bm{\eta}_1\right\}, \\
  p(\bm{\eta}_2 | \tau_2) &\sim \tau_2^{\rank(\bm{F}_2)/2}\exp\left\{-\frac{\tau_2}{2}\bm{\eta}_2^\top \bm{F}_2 \bm{\eta}_2\right\}, \\
  \bm{\epsilon} &\sim N(\bm{0}, \bm{I}/\tau_\epsilon).
\end{align}

If $\bm{W}_1$ and $\bm{W}_2$ have orthonormal columns such that ${\cal C}(\bm{X})$, ${\cal C}(\bm{W}_1)$ and ${\cal C}(\bm{W}_2)$ are pairwise orthogonal, $\bm{F}_1$ and $\bm{F}_2$ are positive definite symmetric matrices, $\tau_j \sim \text{gamma}(a_j, b_j)$ for $j=1,2$, $\tau_\epsilon \sim gamma(a_\epsilon, b_\epsilon)$ and 
\begin{equation}\frac{\mathrm{E}[r_1|\bm{y}]/b_1 + \mathrm{E}[r_2|\bm{y}]/b_2}{\mathrm{E}[\tau_1]/b_1 + \mathrm{E}[\tau_2]/b_2} \leq E[\sigma^2_{\epsilon, NN}|\bm{y}], \end{equation}
where $r_j = \tau_j/\tau_\epsilon$ for $j=1,2$, then $\mathrm{Var}(\delta_\ell|\bm{y}) \leq \mathrm{Var}(\delta_{NN,\ell}|\bm{y})$ for $\ell=1,\dots,p$.

\label{thm:network-2randomeffects-2b}
\end{theorem}

The proof of this theorem is provided in Appendix \plaintextref{A} of the Supplementary Material \citep{articlesupplement}. This theorem shows that under the specified conditions, a model with two random effects restricted to be orthogonal to the covariates yields posterior variances that do not meet the conditions in Definition \ref{def:alleviates-network-confounding}. Applying Theorem \ref{thm:network-2randomeffects-2b} to Restricted Network Regression with $\bm{W}_1=(\bm{I}-\bm{P}_X)\bm{A}$, $\bm{W}_2=(\bm{I}-\bm{P}_X)\bm{B}$, and $\bm{F}_1=\bm{F}_2=\bm{I}_n$ shows that Restricted Network Regression with a continuous response does not meet the conditions in Definition \ref{def:alleviates-network-confounding}.

\citet{khan2020restricted} prove similar theorems for restricted spatial regression models of the form,
\begin{align}
    \bm{y} &= \bm{X}\bm{\delta} + \bm{W}\bm{\eta} + \bm{\varepsilon}, \label{eqn:khan-general-rsr-start}\\
    p(\bm{\eta} | \tau_s) &\propto \tau_s^{\mathrm{rank}(\bm{F})/2}\exp\left\{-\frac{\tau_s}{2}\bm{\eta}^\top \bm{F} \bm{\eta}\right\}  \\
    \bm{\varepsilon} &\sim N(\bm{0}, \bm{I}/\tau_\varepsilon) \label{eqn:khan-general-rsr-end}, 
\end{align}
observing that that the model form in (\ref{eqn:khan-general-rsr-start})-(\ref{eqn:khan-general-rsr-end}) encompasses the ICAR model, the non-spatial model, and restricted spatial regression models from \citet{reich2006effects}, \citet{hughes2013dimension}, and \citet{prates2019alleviating}. The model in (\ref{eqn:khan-general-rsr-start})-(\ref{eqn:khan-general-rsr-end}) is ``restricted'' if ${\cal C}(\bm{X})$ and ${\cal C}(\bm{W})$ are orthogonal. In fact, this form also encompasses continuous network models that include one additive sender or one additive receiver random effect, but not both, motivating the need for Theorem \ref{thm:network-2randomeffects-2b}.

\section{Simulation Study} \label{sec:simulation-study}

In this section we investigate properties of Restricted Network Regression through simulations. We confirm the theoretical results from Section \ref{sec:rsr-theory} using continuous network data and show that neither continuous nor binary Restricted Network Regression alleviate network confounding. However, we show that both mitigate network confounding relative to non-restricted network regression and network regression with no random effects.
All of the following simulations involve data with varying levels of excess nodal variation, using models with both sender/receiver covariates and sender/receiver random effects. 

In addition to evaluating Restricted Network Regression, we chose to evaluate choices of prior distribution on $\sigma^2_a$ and $\sigma^2_b$. A common choice is the inverse-gamma distribution, e.g., in the \texttt{amen} package \citep{hoff2020amenpkg}. However, the half-Cauchy distribution, $\sigma_a \sim \mathrm{Cauchy}^+(0,1)$, is a less informative distribution recommended by \citet{Gelman2006} for random effect variances in hierarchical models. We compare five models:
\begin{enumerate}[align=left]
    \item[(NoRE)] Network model with no random effects,
    \item[(NR.ig)] Network model with additive random effects and inverse-gamma priors,
    \item[(NR.hc)] Network model with additive random effects and half-Cauchy priors,
    \item[(RNR.ig)] Network model with restricted additive random effects and inverse-gamma priors,
    \item[(RNR.hc)] Network model with restricted additive random effects and half-Cauchy priors.
\end{enumerate}
We show that Restricted Network Regression mitigates network confounding by providing estimates of $\bm{\delta}$ with lower bias and properly calibrated credible intervals.

\subsection{Simulation 1: Continuous Network Data} \label{sec:simulation-continuous}

Continuous network data were simulated from the network model in (\ref{eqn:network-model-unrestricted-start})-(\ref{eqn:network-model-unrestricted-end}), using the identity function, $g(z_{ij})=z_{ij}$. The design matrix $\bm{X}$ contained an intercept, one sender covariate, one receiver covariate, and one dyadic covariate whose values were drawn independently from a standard normal distribution. Unobserved excess nodal variation (simulated values of $a_i$ and $b_j$, denoted $\bm{a}^*$ and $\bm{b}^*$) was then simulated from one of seven possible scenarios, which varied in the magnitude of the nodal variation and the degree of collinearity between the nodal variation and observed covariates. We control and quantify this latter degree of collinearity using the canonical correlation ($\mathrm{ccor}$) between $\bm{A}\bm{a}^*+\bm{B}\bm{b}^*$ and $\bm{X}$. 
\begin{enumerate}[align=left]
    \item[(G1)] No excess variation: $\bm{a}^* = \bm{b}^* = \bm{0}$,
    \item[(G2)] Small magnitude, no correlation: \\ $\bm{a}^* \sim \text{Normal}(0, 0.25)$, $\bm{b}^* \sim \text{Normal}(0, 0.25)$, $\mathrm{ccor}(\bm{A}\bm{a}^* + \bm{B}\bm{b}^*, \bm{X}) = 0$,
    \item[(G3)] Small magnitude, slight correlation: \\ $\bm{a}^* \sim \text{Normal}(0, 0.25)$, $\bm{b}^* \sim \text{Normal}(0, 0.25)$, $\mathrm{ccor}(\bm{A}\bm{a}^* + \bm{B}\bm{b}^*, \bm{X}) = 0.1$,
    \item[(G4)] Small magnitude, strong correlation: \\ $\bm{a}^* \sim \text{Normal}(0, 0.25)$, $\bm{b}^* \sim \text{Normal}(0, 0.25)$, $\mathrm{ccor}(\bm{A}\bm{a}^* + \bm{B}\bm{b}^*, \bm{X}) = 0.9$,
    \item[(G5)] Large magnitude, no correlation: \\$\bm{a}^* \sim \text{Normal}(0, 1)$, $\bm{b}^* \sim \text{Normal}(0, 1)$, $\mathrm{ccor}(\bm{A}\bm{a}^* + \bm{B}\bm{b}^*, \bm{X}) = 0$,
    \item[(G6)] Large magnitude, slight correlation: \\$\bm{a}^* \sim \text{Normal}(0, 1)$, $\bm{b}^* \sim \text{Normal}(0, 1)$, $\mathrm{ccor}(\bm{A}\bm{a}^* + \bm{B}\bm{b}^*, \bm{X}) = 0.1$,
    \item[(G7)] Large magnitude, strong correlation: \\$\bm{a}^* \sim \text{Normal}(0, 1)$, $\bm{b}^* \sim \text{Normal}(0, 1)$, $\mathrm{ccor}(\bm{A}\bm{a}^* + \bm{B}\bm{b}^*, \bm{X}) = 0.9$.
\end{enumerate}

In all scenarios, the components of $\bm{a}^*$ and $\bm{b}^*$ were first generated i.i.d., then the vectors were projected to have the desired canonical correlation according to the algorithm in Appendix \plaintextref{B} of the Supplementary Material \citep{articlesupplement}. Scenarios G2, G3, and G4 were chosen to represent nodal variation smaller than the effect of one covariate, while scenarios G5, G6, and G7 were chosen to represent variation comparable to the effect of one covariate. The slight correlation scenarios (G3 and G6) represent situations with subjectively low correlation between nodal variation and covariates, while the strong correlation scenarios (G4 and G7) represent situations with subjectively high correlation between nodal variation and covariates.
For each of the seven scenarios, 100 values of $\bm{X}$, $\bm{a}^*$, and $\bm{b}^*$ were generated. Then for each set of covariates and random effects, 200 values of the random error $\bm{\epsilon}$ were generated each from a standard normal distribution ($\sigma^2_\epsilon = 1$), resulting in 20,000 simulated data sets for each scenario.

\begin{figure}
\begin{tabular}{m{0.04\linewidth}m{0.90\linewidth}}
(A) & \includegraphics[width=0.9\linewidth]{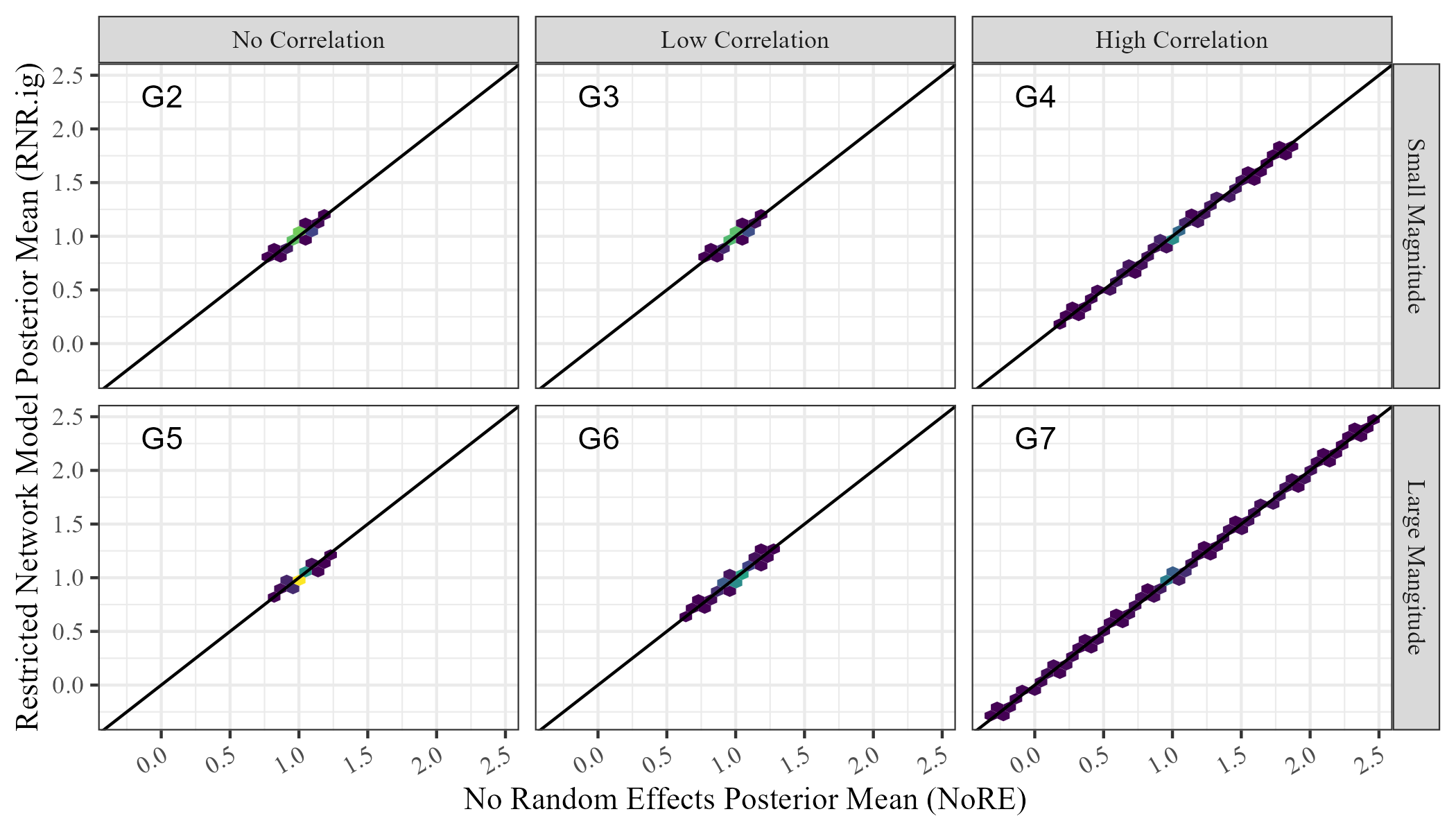} \\
(B) & \includegraphics[width=0.9\linewidth]{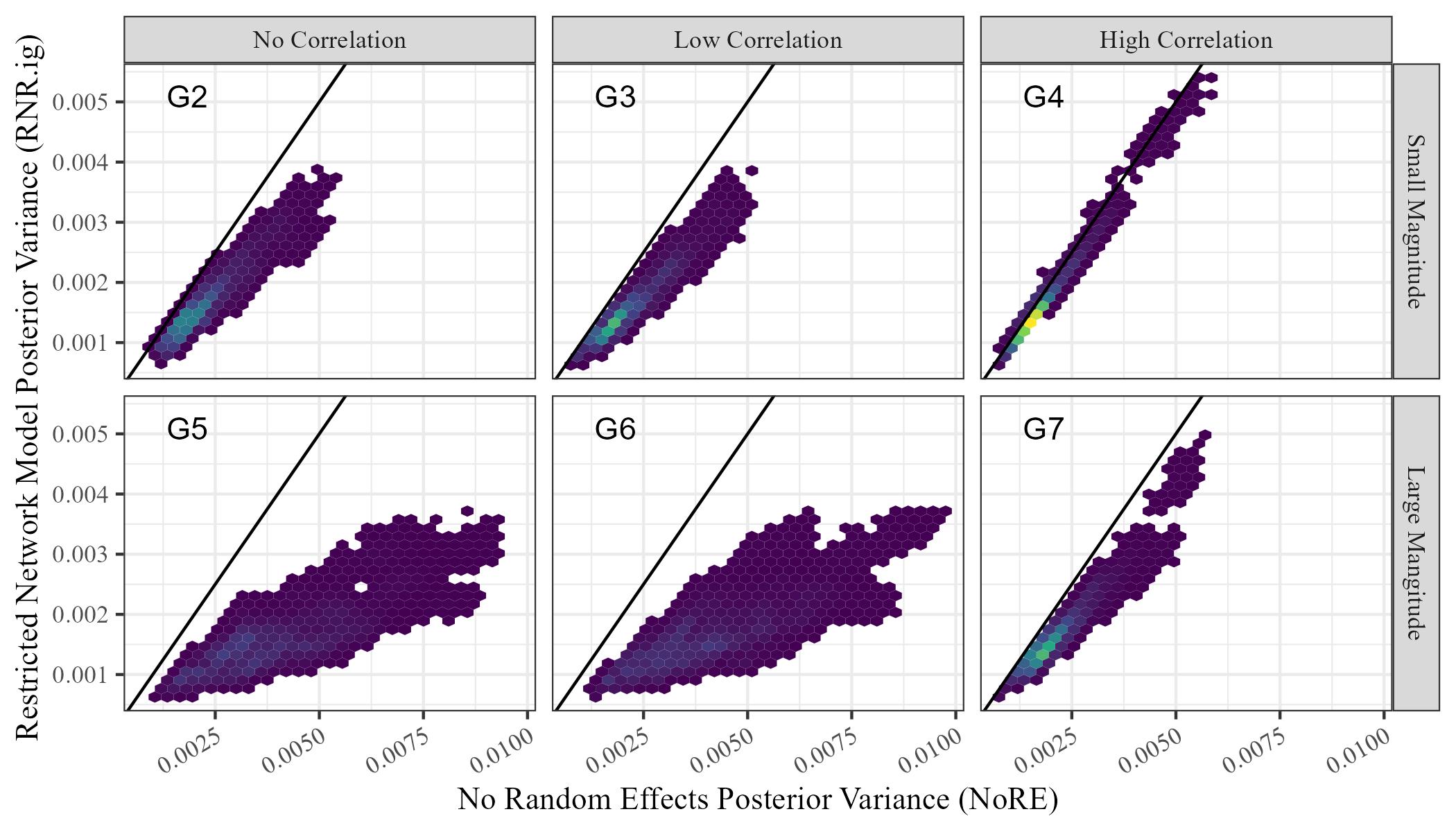}
\end{tabular}
\caption{Comparison of posterior means (A) and variances (B) for a receiver covariate using a network regression model with no random effects (NoRE) and a Restricted Network Regression model (RNR.ig) with a continuous response. Each panel is one of scenarios G2 through G7. Each panel shows a heatmap of the $100 \times 200$ simulated data sets for each scenario, and the $y=x$ line is drawn on each panel, which represents an equal value from both models. The area below the line contains smaller values in the Restricted Network Regression model, and the area above the line contains larger values in the Restricted Network Regression model.}
\label{figure-simulation-continuous-meanvar}
\end{figure}

We compare the posterior means and posterior variances of the unconditional regression parameters in NoRE and RNR.ig. Figure \ref{figure-simulation-continuous-meanvar} shows these values for the receiver covariate, $\delta_r$. We see that the posterior means are equal for both models on all simulated data sets, and the posterior variance in the Restricted Network Regression model (RNR.ig) is less than or equal to the posterior variance in the model with no random effects (NoRE). Similar results were observed for the sender covariate. The equality of posterior means and this observed inequality of posterior variances validate the result of Theorem \ref{thm:network-2randomeffects-2b} empirically, and show that continuous Restricted Network Regression does not alleviate network confounding according to Definition \ref{def:alleviates-network-confounding}.

\begin{figure}
    \begin{tabular}{m{0.04\linewidth}m{0.90\linewidth}}
    (A) & \includegraphics[width=0.9\linewidth]{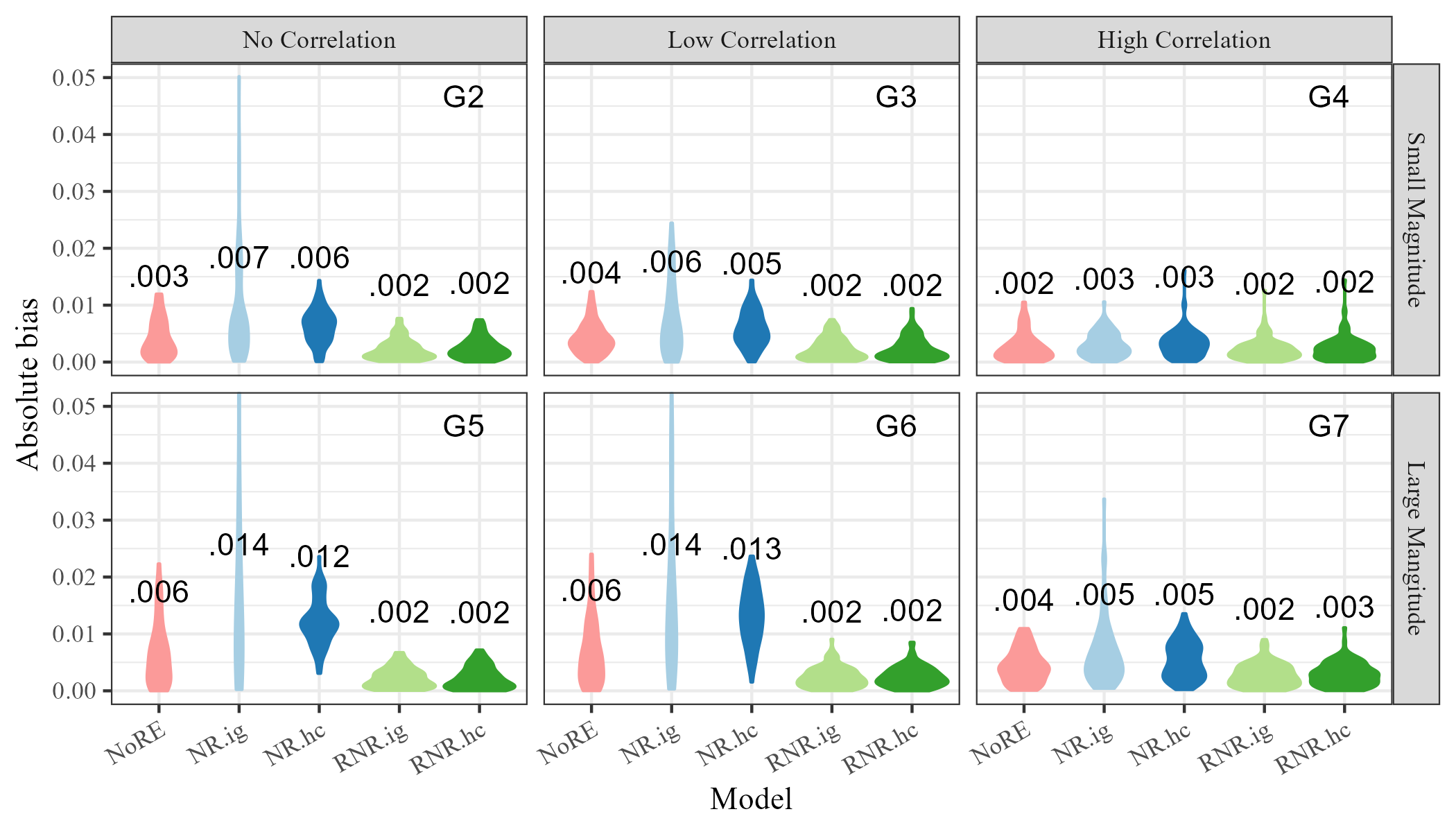} \\
    (B) & \includegraphics[width=0.9\linewidth]{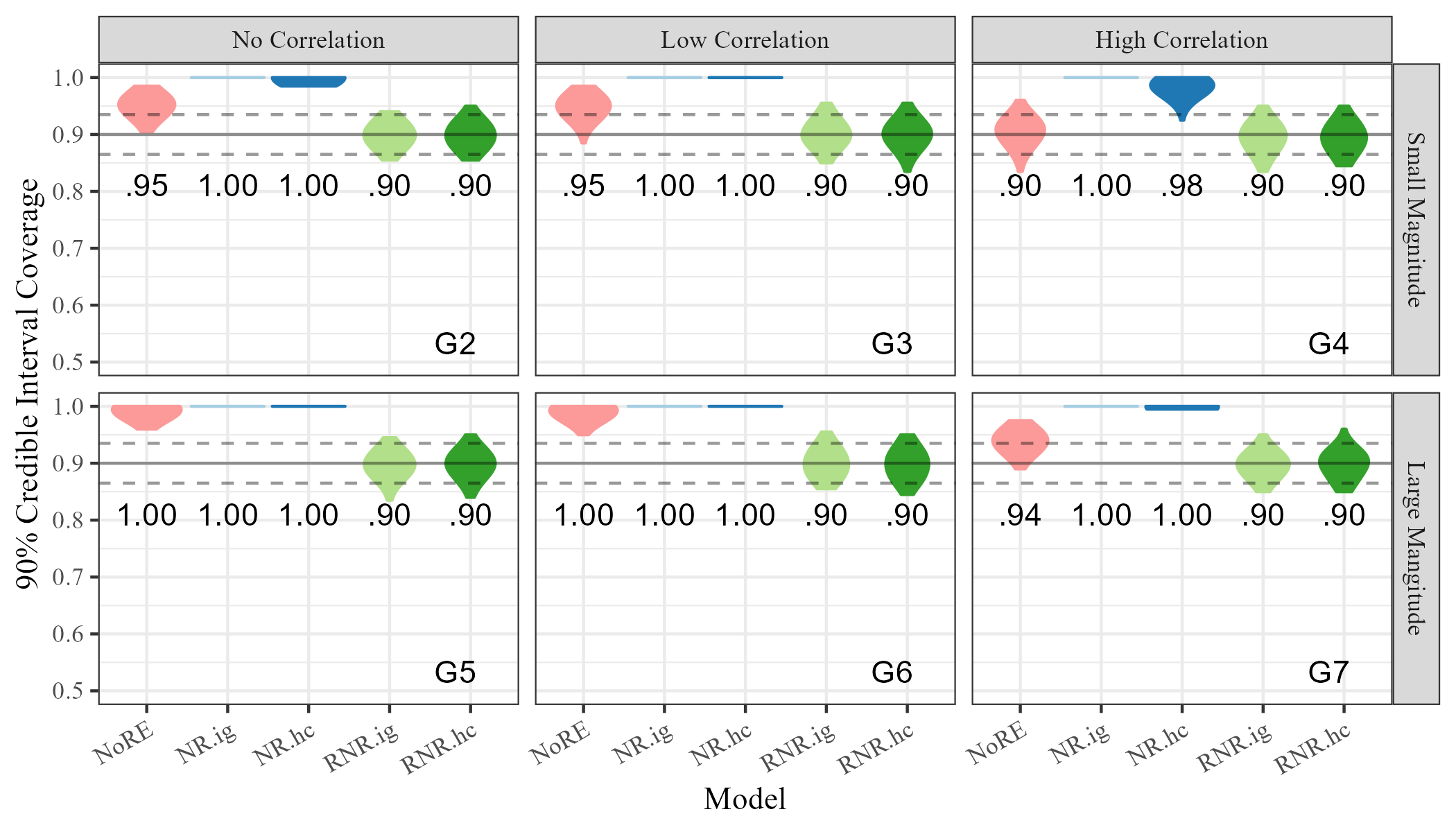} \\
    \end{tabular}
    \caption{Absolute bias (A) and credible interval coverage (B) for all models when estimating $\delta_r$ with continuous data. Median bias and coverage values are printed for each model. Each panel is one of scenarios G2 through G7. Violin plots show the distribution of the estimated absolute bias and coverage values of the 100 simulated values of $\bm{X}$, $\bm{a}^*$, and $\bm{b}^*$. Solid horizontal lines indicate the nominal coverage (90\%) and dashed horizontal lines at 86.5\% and 93.5\% indicate bounds within which the average coverage of a 90\% credible interval should fall over 200 trials.}
    \label{fig:simulation-continuous-coverage}
\end{figure}

To investigate the ability of Restricted Network Regression to mitigate network confounding relative to a model with no random effects and models with non-restricted random effects, we compare bias and coverage of posterior credible intervals for across all models (NoRE through RNR.hc). For each of the 200 trials with each of the 100 values of $\bm{X}$, $\bm{a}^*$, and $\bm{b}^*$, we recorded the difference between the posterior means of $\bm{\delta}$ and and the value $\bm{\delta}^*$ used to generate the data. 
 We also recorded whether the 90\% credible for $\bm{\delta}$ captures $\bm{\delta}^*$. Finally we calculated the average 
 bias across the 200 values of $\bm{y}$ generated for each of the 100 values of $X$, and the proportion of the 200 trials for which $\bm{\delta}^*$ was captured by the credible interval. 

Figure \ref{fig:simulation-continuous-coverage} shows the distribution of the absolute bias and credible interval coverage for $\delta_r$ using each model in each scenario, G2 through G7. Bias appears lower for the restricted models (RNR.ig and RNR.hc) than for other models in scenarios G2, G3, G5, G6, and G7, and approximately equal in G4. The most noticeable difference between the Restricted Network Regression models and others is in credible interval coverage, where models RNR.ig and RNR.hc appear properly calibrated and NoRE, NR.ig, and NR.hc have coverage that is too high. Together, these results suggest the continuous Restricted Network Regression mitigates network confounding relative to both non-restricted network regression and network regression with no random effects.

\subsection{Simulation 2: Binary Network Data} \label{sec:simulation-binary}

Data for the binary network models were simulated in the same way as for the continuous network model (Section \ref{sec:simulation-continuous}), using scenarios G1 through G7, but with the indicator function $g(z_{ij}) = I(z_{ij} > 0)$ to convert latent continuous responses to binary responses. All models were fit to each set of simulated data. We similarly assessed the results of these simulations by comparing posterior means and posterior variances of the unconditional regression parameters in models NoRE and RNR.ig.

\begin{figure}
\begin{tabular}{m{0.04\linewidth}m{0.90\linewidth}}
(A) & \includegraphics[width=0.9\linewidth]{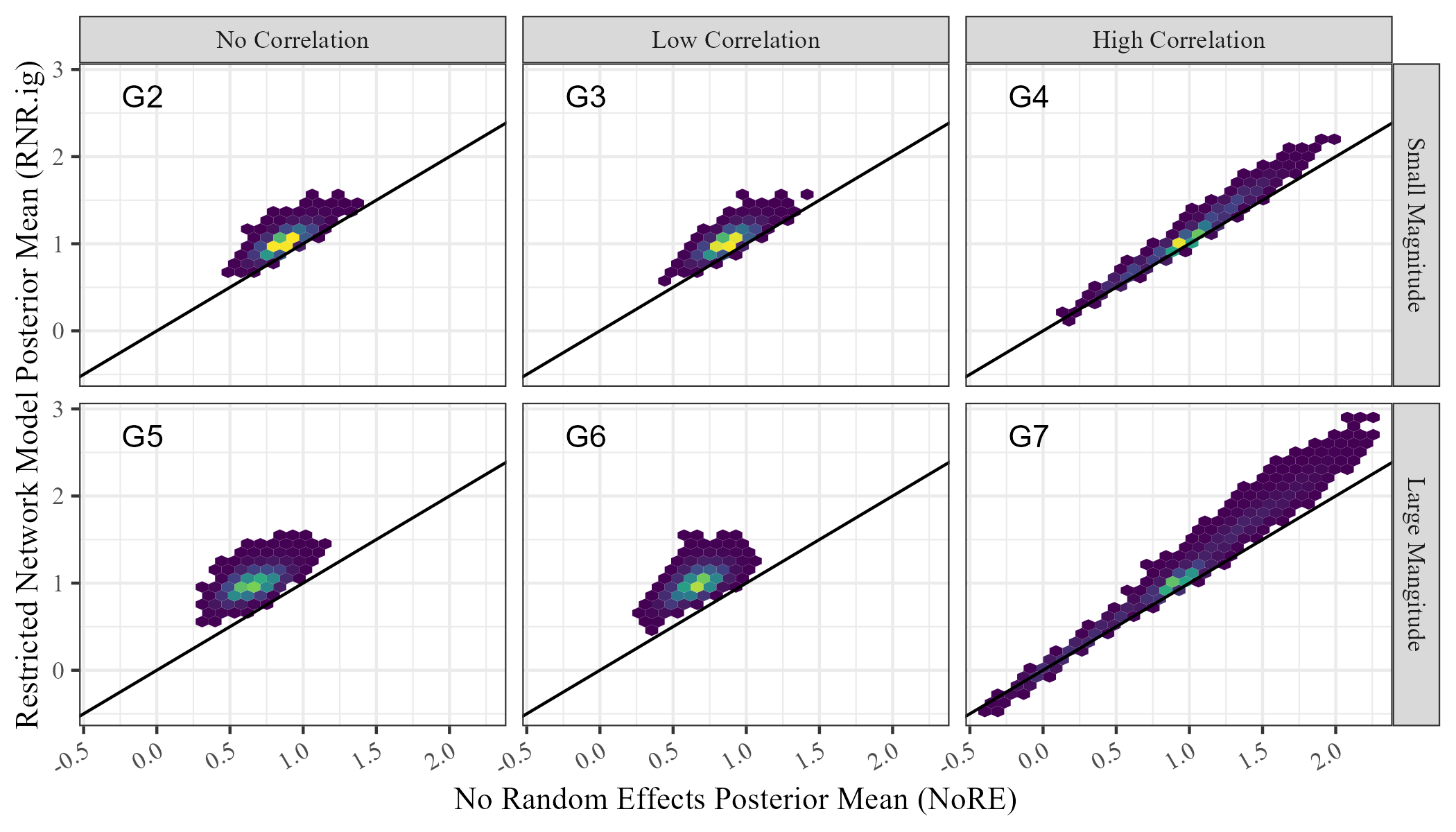} \\
(B) & \includegraphics[width=0.9\linewidth]{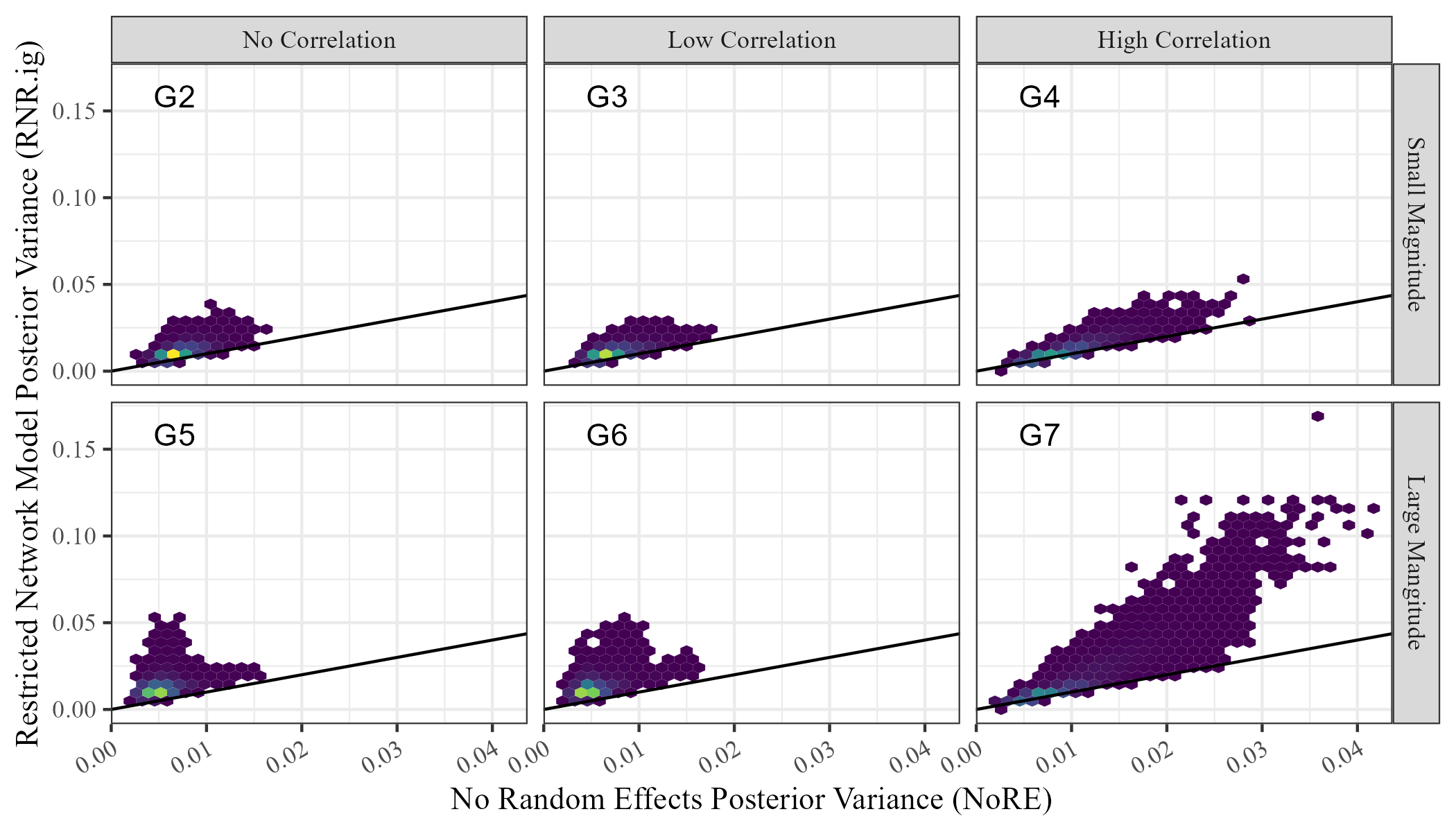} 
\end{tabular}
\caption{Comparison of posterior means (A) and variances (B) for a receiver covariate using a network regression model with no random effects (NoRE) and a Restricted Network Regression model (RNR.ig) with a binary response using a probit link. Each panel is one of scenarios G2 through G7. Each panel shows a heatmap of the $100 \times 200$ simulated data sets for each scenario, and the $y=x$ line is drawn on each panel, which represents an equal value from both models. The area below the line contains smaller values in the Restricted Network Regression model, and the area above the line contains larger values in the Restricted Network Regression model.}
\label{figure-simulation-binary-meanvar}
\end{figure}

Figure \ref{figure-simulation-binary-meanvar} shows the comparison between posterior means and variances of $\delta_r$ for models NoRE and RNR.ig with binary data. This comparison is notably different from the comparison for continuous data in Figure \ref{figure-simulation-continuous-meanvar}. First, the posterior means produced by each model are not equal. Second, the posterior variance of the regression coefficient in the Restricted Network Regression model is now greater than in the network model with no random effects. This demonstrates empirically that the implications of Theorem \ref{thm:network-2randomeffects-1} and Theorem \ref{thm:network-2randomeffects-2b} do not apply to models with non-Gaussian responses due to the inequality of posterior means. However, this inequality of means also demonstrates that probit Restricted Network Regression also does not alleviate network confounding.

\begin{figure}
    \begin{tabular}{m{0.04\linewidth}m{0.90\linewidth}}
    (A) & \includegraphics[width=0.9\linewidth]{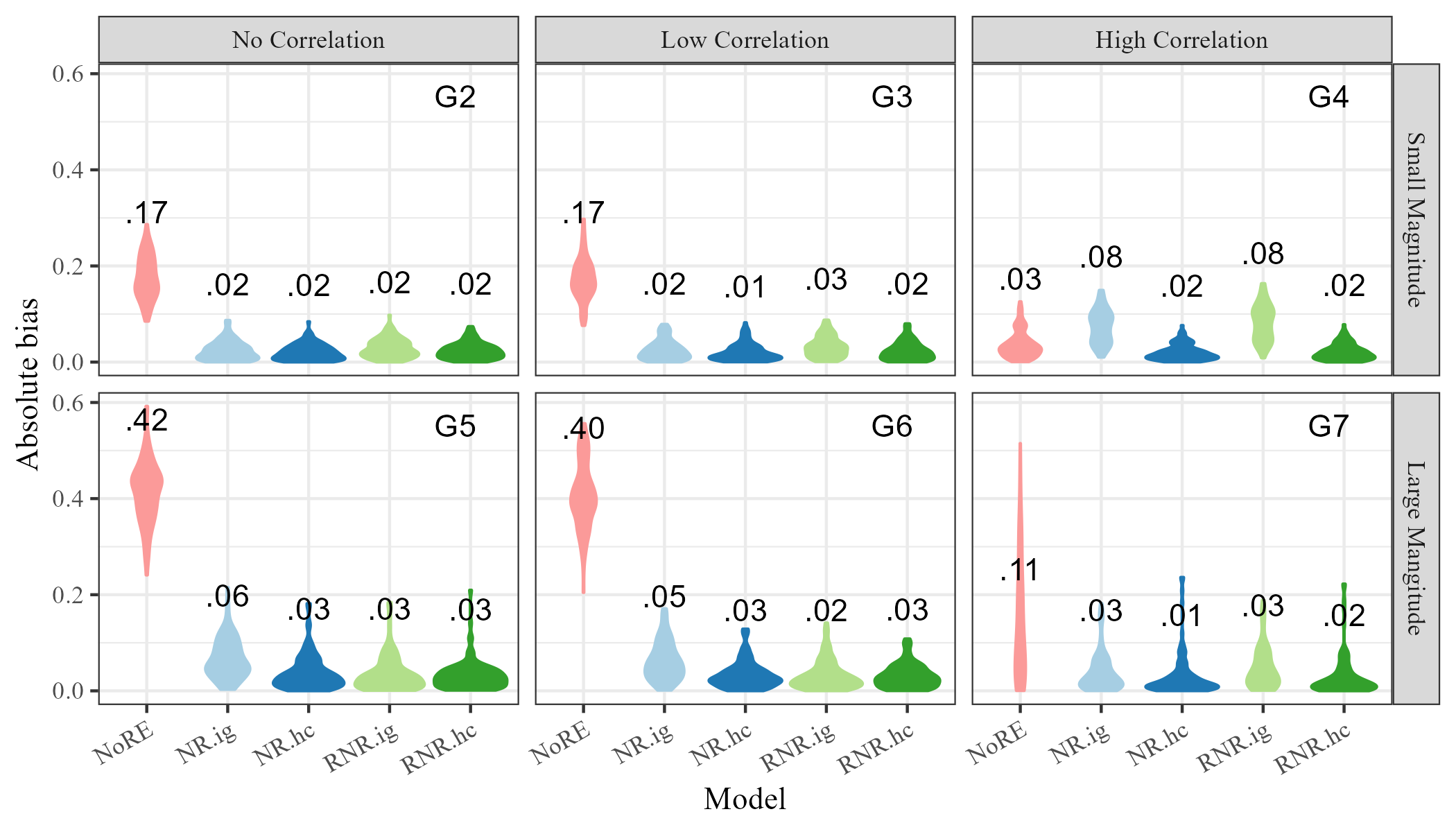} \\
    (B) & \includegraphics[width=0.9\linewidth]{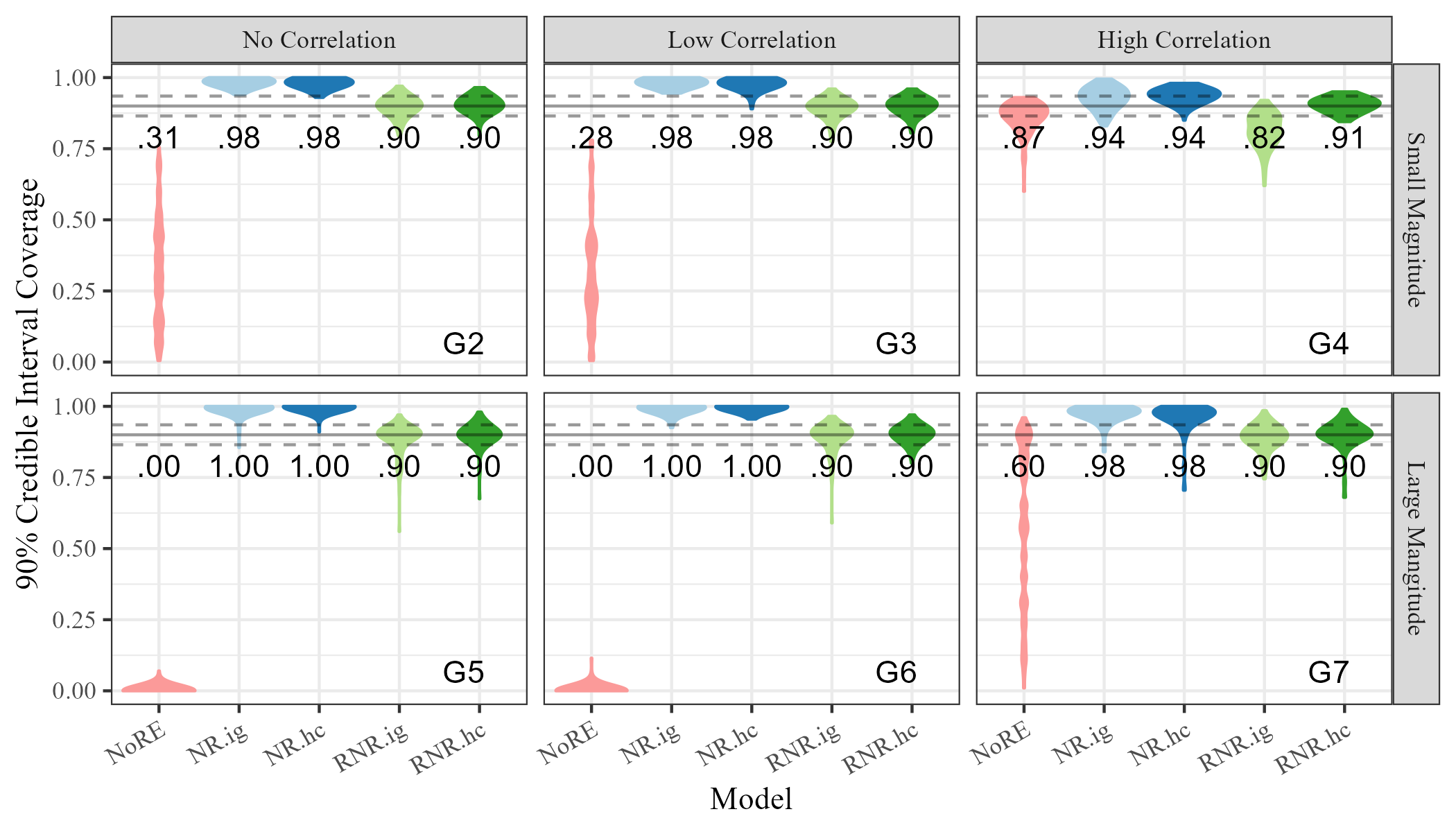} \\
    \end{tabular}
    \caption{Absolute bias (A) and credible interval coverage (B) for all models when estimating $\delta_r$ with binary data. Median absolute bias and coverage values are printed for each model. Each panel is one of scenarios (G2) - (G7). Violin plots show the distribution of the estimated absolute bias and coverage values of the 100 simulated values of $\bm{X}$, $\bm{a}^*$, and $\bm{b}^*$. Solid horizontal lines indicate the nominal coverage (90\%) and dashed horizontal lines at 86.5\% and 93.5\% indicate bounds within which the average coverage of a 90\% credible interval should fall over 200 trials.}
    \label{fig:simulation-binary-coverage}
\end{figure}

Figure \ref{fig:simulation-binary-coverage} shows the absolute bias and coverage estimates for $\delta_r$ in scenarios G2 through G7 using all models. 
The bias for model NoRE is the highest in general, with all other models having approximately equal absolute bias except in scenario G4. In most scenarios, the coverage of model NoRE is significantly lower than 90\%, while the coverage of the Restricted Network Regression model with both inverse-gamma (RNR.ig) and half-Cauchy (RNR.hc) priors is much closer to 90\%. Again, the non-restricted models (NR.ig and NR.hc) have coverage that is higher than 90\% in all scenarios. The Restricted Network Regression model with half-Cauchy random effect priors (RNR.hc) has coverage within the expected range in all scenarios. A notable exception is scenario G4, in which the models with inverse-gamma priors on $\sigma_a$ and $\sigma_b$ (NR.ig and RNR.ig), have higher bias than their half-Cauchy counterparts. The coverage for model RNR.ig is also noticeably lower than RNR.hc. Here, the prior selection affects the model's ability to mitigate network confounding. While model RNR.hc mitigates network confounding relative to NR.hc and RNR.ig mitigates network confounding relative to NR.ig, RNR.ig does not mitigate network confounding relative to NR.hc due to higher bias in this scenario.

\section{Eurovision Voting Network Analysis} \label{sec:eurovision-analysis}

The Eurovision Song Contest is an annual competition in which European countries compete by submitting the best song by an artist from their country. 
The contest culminates in a final round, where the remaining 26 competitors perform their songs for a TV audience. All participating countries then vote for their top ten songs through judges and/or phone-in voting. Points are awarded according to votes (12 points for first, 10 points for second, then 8 through 1 points for third through tenth) and the total determines the winner.

The contest is extremely popular, drawing 182 million viewers in 2019 \citep{eurovision-viewership}. This popularity has meant that the contest has been of interest for study, especially the study of voting patterns \citep[e.g.,][]{GINSBURGH2008eurovision, spierdijk2009structure}. Because competing countries are a subset of voting countries, vote data can naturally be represented as a directed graph with the countries as nodes and an edge from country $i$ to country $j$ representing a top-10 vote by country $i$ for country $j$. Edges may be labeled with ranked votes if desired. Eurovision votes have been analyzed as a network by, for example, \citet{YAIR1995unite}, \citet{FENN2006europe}, and \citet{dangelo2019}. 

\begin{figure}
    \centering
    \includegraphics[width=0.95\linewidth]{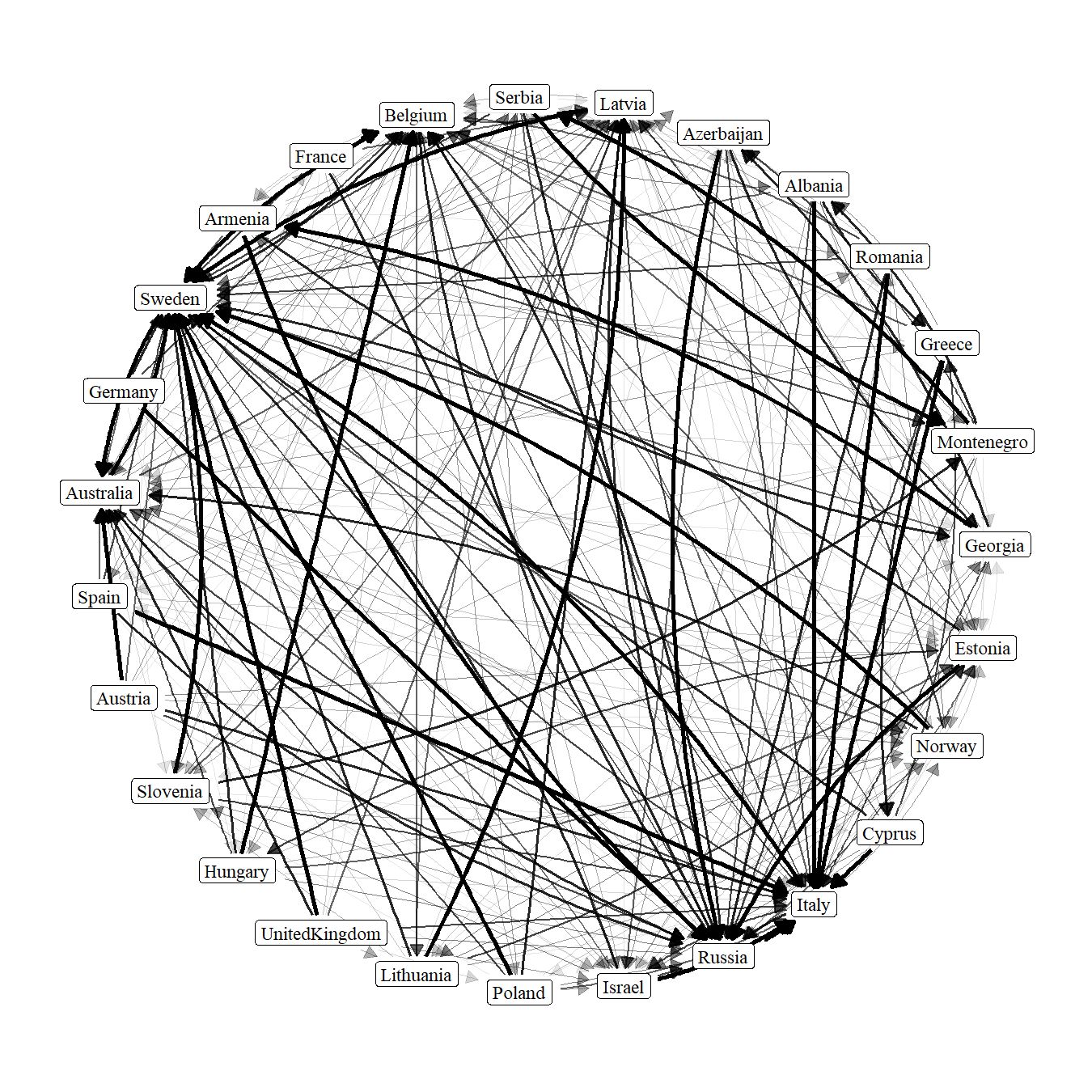}
    \caption{Illustration of vote data from the 2015 Eurovision Song Contest final round as a network. Edges point from voter to song, with darker lines indicating higher rankings. The contest's eventual winner, Sweden, receives a majority of the high-ranked votes.}
    \label{fig:eurovision-vote-network}
\end{figure}

Countries may have other measurable qualities that are related to how well they score in the contest results. For example, countries with larger populations have more musicians from which to select a contestant. A country's wealth may be associated with the reach of its cultural exports, leading to more votes received from its trading partners. The Eurovision Song Contest is also the subject of bets predicting its winner. Betting markets reflect the collective knowledge of their participants, which in this case includes knowledge of the specific songs entered by each country, and it is reasonable to think they may be predictive of the outcome. For example, \citet{spann2009sports} finds betting odds to be predictive of match results in the German premier soccer league. Analyzing the relationship between population, wealth, and betting odds, and the Eurovision contest voting outcome fits a network regression framework naturally. 

We restrict our analysis to the year 2015 and the 27 countries with entries in the final round of that year (Australia, as a new contestant, was given an automatic berth to the final round). 
The response data consist of a vote network represented as a $27 \times 27$ matrix (Figure \ref{fig:eurovision-vote-network}). The receiver covariate data consist of three dimension 27 vectors of song or country attributes: the log median odds from 16 popular European betting sites for each song to win the contest, the log 2015 population of each competing country, and the log 2015 GDP per capita of each competing country. We log-transform the receiver covariates before using them as covariates in a regression model because they are either right-skewed (GDP, population) or because we believe there to be a logarithmic relationship between the predictor and the response (betting odds). We also include a dyadic covariate for country contiguity which was found to be explanatory in \citet{dangelo2019}. Country contiguity is an undirected network represented as a symmetric $27 \times 27$ binary matrix where a 1 indicates that two countries share a border and a 0 indicates otherwise.  Visualizations of these covariates are available in Appendix \plaintextref{C} of the Supplementary Material \citep{articlesupplement}. Votes were represented as ranks (1-10 with 10 the highest). This data is freely available and easily compiled by hand \citep{eurovision-votes, eurovision-betting, CEPII2022gravity, un2015world, worldbank2015gdp}. Any pairs $i,j$ where country $i$ did not vote for country $j$ in its top 10 were coded as zero. Our focus on a single year is due to the fact that both betting odds and column random effects are song-specific, and therefore year-specific. Therefore additional years in this analysis cannot be treated like replicates in the style of \citet{dangelo2019}. We examine the effects of the covariates on Eurovision voting by performing a network regression analysis without random effects, with receiver random effects, and with restricted receiver random effects. We investigate the effect of Restricted Network Regression on regression parameter estimates and interpretation compared to either alternative model.

\subsection{Network Model with No Random Effects} \label{sec:eurovision-non-network}

The base model with no network-structured random effects has the form,
\begin{align}
\bm{y} &= g(\bm{z}), \\
z_{ij} &= \delta_{CC}x^1_{ij} + \delta_{Odds}x^2_j + \delta_{Pop}x^3_j + \delta_{GDP}x^4_j + \varepsilon_{ij}, \label{eqn:eurovision-model-no-random-effects} \\
\varepsilon_{ij} &\overset{\mathrm{i.i.d.}}{\sim} N(0,1).
\end{align}
We used the relative rank likelihood \citep[RRL;][]{pettitt1982inference, hoff_fosdick_volfovsky_stovel_2013}, implemented with a function $g(\cdot)$ which maps continuous values $z_{ij}$ to the observed ranks $y_{ij}$ in the following way: for any voting country $i$ and two entered songs $j$ and $j'$, $y_{ij} > y_{ij'}$ implies $z_{ij} > z_{ij'}$. This imposes no relationship between the responses for different voting countries, so we cannot infer row effects \citep{hoff_fosdick_volfovsky_stovel_2013}.

\begin{figure}
\includegraphics[width=\linewidth]{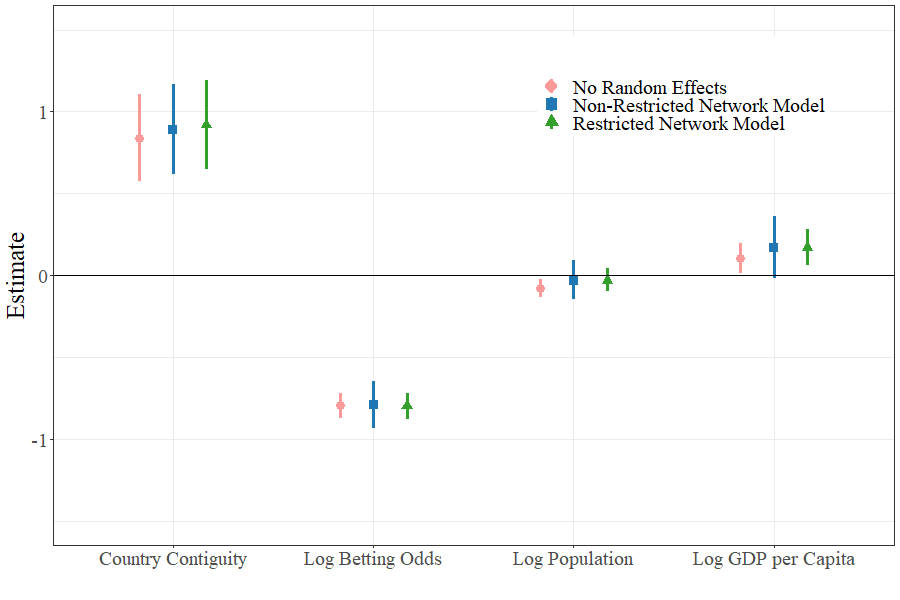}
\caption{Comparison of posterior means and 90\% credible intervals for $\bm{\beta}$ (Non-Restricted Network Model) or $\bm{\delta}$ (No Random Effects and Restricted Network Model). As betting odds, population, and GDP are receiver covariates confounded with the random effects, inference on these regression parameters changes the most between the three models. The posterior distribution for country contiguity, a dyadic covariate, is less affected by the choice of model.}
\label{fig:results-eurovision-ci}
\end{figure}

We analyze and interpret the regression parameter estimates through posterior means and 90\% credible intervals for $\bm{\delta}$ (Figure \ref{fig:results-eurovision-ci}). As these are estimates of $\bm{\delta}$ and there are no random effects in the model, we interpret them as the unconditional effect of the fixed effects on the response. 
All fixed covariates--country contiguity, log betting odds, log GDP per capita, and log population--appear to have an effect on the voting outcomes. Country contiguity has a large positive effect, agreeing with \citet{dangelo2019} that countries are more likely to vote for their neighbors' songs. Log betting odds have a large negative effect, which shows that betting markets are predictive of the Eurovision outcome as larger odds are associated with lower predicted probability of winning. Log GDP per capita has a small positive effect, indicating that wealthier countries are more likely to receive votes than less wealthy countries. Log population has a small negative effect. All other things being equal, more populous countries are less likely to receive votes than less populous countries.

\subsection{Non-Restricted Network Model} \label{sec:eurovision-non-restricted}

We fit a network model with a receiver random effect ($b_j$) to account for song heterogeneity not explained by the betting odds, population, or GDP. 
This network model has the form,
\begin{align}
\bm{y} &= g(\bm{z}), \\
z_{ij} &= \beta_{CC}x^1_{ij} + \beta_{Odds}x^2_j + \beta_{Pop}x^3_j + \beta_{GDP}x^4_j + b_j + \varepsilon_{ij}, \label{eqn:eurovision-model-unrestricted} \\
b_j &\overset{\mathrm{i.i.d.}}{\sim} N(0, \sigma^2_b), \\
\varepsilon_{ij} &\overset{\mathrm{i.i.d.}}{\sim} N(0, 1).
\end{align}
Leaving the random effects non-restricted is appropriate depending on the intended interpretation of the terms in the model. For example, restricted regression is not recommended in the case of ``Scheff\'e-style'' random effects: random effects whose values are considered as draws from a population which is of interest, even though the values of the effects themselves are not \citep{Hodges2010}. If we consider the population of countries to be all those eligible to participate in the contest, or all those who competed in the initial rounds, then the selection of countries in the final round is only a subset of the population. If the primary interest is studying the population of all eligible countries rather than the propensity of individual countries to receive votes, the receiver random effects could be considered ``Scheff\'e-style'' and restricted regression may not be an appropriate choice. 

Because the model contains receiver random effects, the regression effects represent the effect of the covariates on the response conditioned on $\bm{b}$. 
The posterior means and credible intervals of the regression parameters in this model are noticeably different than in the network model without random effects (Figure \ref{fig:results-eurovision-ci}). In this case, the country contiguity retains its large positive effect and log betting odds retains its large negative effect. We notice that the width of the 90\% posterior credible intervals for $\beta_{Odds}$, $\beta_{GDP}$, and $\beta_{Pop}$ are wider than the credible intervals for $\delta_{Odds}$, $\delta_{GDP}$, and $\delta_{Pop}$ from the model in (\ref{eqn:eurovision-model-no-random-effects}), indicating greater uncertainty about the conditional effect of these covariates than the unconditional effect. The estimate of $\beta_{Pop}$ is also smaller in magnitude and has a credible interval which includes zero. The changes in credible interval width and posterior mean illustrate the impact of network confounding.  

\begin{table}
    \caption{Comparison of covariate effect estimates from Restricted Network Regression to other models. For each covariate and each comparison model, the ratio of posterior means and posterior credible interval widths are shown. Numbers smaller than 1 indicate smaller posterior means or narrower credible intervals in the Restricted Network Regression model.}
    \label{table:eurovision-results-comparison}
    \centering
\begin{tabular}{lllll}
\toprule
\multicolumn{1}{c}{ } & \multicolumn{4}{c}{Comparison Model} \\
\cmidrule(l{3pt}r{3pt}){2-5}
\multicolumn{1}{c}{ } & \multicolumn{2}{c}{No Random Effects} & \multicolumn{2}{c}{Non-Restricted Model} \\
\cmidrule(l{3pt}r{3pt}){2-3} \cmidrule(l{3pt}r{3pt}){4-5}
Covariate & Mean Ratio & Width Ratio & Mean Ratio & Width Ratio\\
\midrule
Log Betting Odds & 1.004 & 1.045 & 1.008 & 0.539\\
Log Population & 0.389 & 1.247 & 1.000 & 0.598\\
Log GDP per Capita & 1.656 & 1.173 & 1.004 & 0.566\\
Country Contiguity & 1.091 & 1.031 & 1.032 & 0.997\\
\bottomrule
\end{tabular}
\end{table}

\subsection{Restricted Network Model} \label{sec:eurovision-restricted}

We fit a network model with a receiver random effect ($\bm{b}$) to account for song heterogeneity not explained by the column covariates, projected to be orthogonal to the fixed effects of betting odds, population, or GDP. Specifically, we set
\begin{align}
    \bm{y} &= g(\bm{z}), \\
    \bm{z} &= \bm{X}\bm{\delta} + (\bm{I} - \bm{P}_X)\bm{B}\bm{b} + \bm{\varepsilon}.
\end{align}
Since the association between $\bm{y}$ and $\bm{X}$ is of primary interest, we would like estimates of $\bm{\delta}$ instead of $\bm{\beta}$. If we do not want to infer about the population of countries which did not compete in this year's final round, then the random effects in the model constitute the entire population of interest. 
Since in this analysis we are using data only from the final round of the 2015 contest, restricted regression would be appropriate. In the Restricted Network Regression model, the regression effects once again represent the unconditional effect of the covariates on the response as the collinearity with the random effect has been removed.

Table \ref{table:eurovision-results-comparison} compares the magnitude of the regression parameter estimates and their credible interval widths from the Restricted Network Regression model to those in the other models. Compared to the model with no random effects, the receiver effects for log population and log GDP per capita exhibit noticeably different posterior means and larger posterior credible intervals, while the effect for log betting odds shows approximately equal posterior mean and only slightly larger credible interval (Figure \ref{fig:results-eurovision-ci}). These results mirror what was observed with the binary data in the earlier simulation study. Based on that study, we expect the estimates from Restricted Network Regression to more accurately capture the unconditional effect of the covariates on the response. Compared to the non-restricted network model, all regression parameters have approximately equal posterior means and the receiver covariate effects have smaller posterior credible intervals. Restricted Network Regression allows the excess network-structured variation in $\bm{y}$ to be accounted for via the random effects $b_j$, while avoiding the network confounding and inflated standard errors in the non-restricted network model.

\section{Discussion} \label{sec:discussion}

In this paper, we introduced Restricted Network Regression for models with additive network random effects and established its connection to restricted spatial regression. We characterized the network confounding of the network regression model with additive random effects and node-level covariates, which Restricted Network Regression addresses by forcing the column spaces of the fixed and random effects to be mutually orthogonal. We provided conditions for network regression models to alleviate and mitigate network confounding, and proved that Restricted Network Regression does not alleviate network confounding with theoretical results and through simulation. However, we showed through simulation that Restricted Network Regression does mitigate network confounding relative to network regression with no random effects and non-restricted network regression with continuous and binary response data. Restricted Network Regression produces less bias and properly calibrated credible intervals for regression parameters relative to network regression without random effects and non-restricted network regression.

We also explored through simulation the effect of a half-Cauchy prior on the variance components of the network random effects. We found that this prior resulted in comparable bias and credible interval coverage for the unconditional regression effects to the conjugate inverse-gamma prior in probit Restricted Network Regression, with better bias and credible interval coverage in some scenarios.

Finally, we applied Restricted Network Regression to a dataset of Eurovision Song Contest voting. We interpreted the model estimates of Restricted Network Regression alongside those from a model without network-structured random effects and those from a non-restricted network regression model. For the three receiver covariates in the model, the choice of model affected both their posterior mean point estimates and their credible interval estimates. The change in credible interval width is noticeable for all three receiver covariates. Uncertainty, as indicated by the width of posterior credible intervals, increases after adding random effects to the model, but decreases again after restricting them. The widths of the credible intervals for the receiver covariates with restricted random effects is larger than without random effects but smaller than with non-restricted random effects.

Future work in this area includes developing Restricted Network Regression for other forms of network random effects such as multiplicative effects \citep{hoff2021additive} or latent space distance effects \citep{hoffetal2002}. It is less clear which covariates may be confounded with such effects, and whether restricting these random effects can have the same benefits as with additive effects and non-Gaussian data. Theoretical results for binary or other non-Gaussian data that describe the posterior distribution are also needed to make stronger conclusions about Restricted Network Regression on non-Gaussian data. Restricted network regression can also be expanded to include bipartite network data or longitudinal network data \citep[e.g.,][]{marrs2020inferring}.




\section*{Supplement}
\textbf{Supplement to ``Restricted Regression in Networks"}
    
This supplementary document contains proofs of theorems in this paper, additional theorems not presented in this paper, details of simulation studies and details of data analysis.

\section*{Acknowledgements}
K. Keller acknowledges the support of NSF Grant 1856229 for this work.

\bibliographystyle{apalike}
\bibliography{rnr.bib}

\end{document}


\def\spacingset#1{\renewcommand{\baselinestretch}%
{#1}\small\normalsize} \spacingset{1}


{
  \title{\bf Supplement to ``Restricted Regression in Networks"}
  \author{Ian Taylor\\
    Department of Statistics, Colorado State University\\
    and \\
    Kayleigh P. Keller \\
    Department of Statistics, Colorado State University\\
    and \\
    Bailey K. Fosdick\thanks{
    bailey.fosdick@cuanschutz.edu}\hspace{.2cm}\\
    Department of Biostatistics and Informatics, Colorado School of Public Health
    }
  \maketitle
}

\newpage
\spacingset{1.45}

\appendix

\section{Theoretical Results} \label{apx:proofs-of-theorems}

\subsection{Proof of Theorem \plaintextref{1}}

We use the laws of total expectation and variance. Starting with the full conditional distribution of $\bm{\delta}$,
$$\bm{\delta} | \bm{a}, \bm{b}, \sigma^2_a, \sigma^2_b, \sigma^2_\varepsilon, \bm{y} \sim N\left((\bm{X}^\top \bm{X})^{-1} \bm{X}^\top \bm{y}, \sigma^2_\varepsilon (\bm{X}^\top \bm{X})^{-1}\right),$$
we have
\begin{align}
    \mathrm{E}[\bm{\delta}|\bm{y}] &= \mathrm{E}[\mathrm{E}[\bm{\delta}|\bm{a}, \bm{b}, \sigma^2_a, \sigma^2_b, \sigma^2_\varepsilon, \bm{y}]|\bm{y}] = (\bm{X}^\top \bm{X})^{-1} \bm{X}^\top \bm{y} \\
    \mathrm{Var}(\bm{\delta} | \bm{y}) &= \mathrm{E}[\mathrm{Var}(\bm{\delta}|\bm{a}, \bm{b}, \sigma^2_a, \sigma^2_b, \sigma^2_\varepsilon, \bm{y})|
    \bm{y}] + \mathrm{Var}(\mathrm{E}[\bm{\delta}|\bm{a}, \bm{b}, \sigma^2_a, \sigma^2_b, \sigma^2_\varepsilon, \bm{y}]|\bm{y}) \\
    &= \mathrm{E}[\sigma^2_\varepsilon (\bm{X}^\top \bm{X})^{-1}|
    \bm{y}] \\
    &= (\bm{X}^\top \bm{X})^{-1} \mathrm{E}[\sigma^2_\varepsilon | \bm{y}].
\end{align}

\sloppy The above identities depend on having a proper joint posterior distribution $f(\bm{\delta}, \bm{a}, \bm{b}, \sigma^2_a, \sigma^2_b, \sigma^2_\varepsilon | \bm{y})$ and a finite posterior expectation $\mathrm{E}[\sigma^2_\varepsilon | \bm{y}] < \infty$ which are both true assuming proper priors for $\sigma^2_a$, $\sigma^2_b$ and $\sigma^2_\varepsilon$ with a finite prior mean $\mathrm{E}[\sigma^2_\varepsilon] < \infty$.

\qedsymbol

\subsection{Proof of Theorem \plaintextref{2}}

We first simplify the model by combining the random effects as follows:
\begin{align}
\bm{\eta}_\ast &:= \begin{bmatrix} \bm{\eta}_1 \\ \bm{\eta}_2 \end{bmatrix} \\
\bm{W}_\ast &:= \begin{bmatrix} \bm{W}_1 & \bm{W}_2 \end{bmatrix} \\
\bm{F}_\ast &:= \bm{F}_\ast(r_1, r_2) = \begin{bmatrix} r_1 \bm{F}_1 & 0 \\ 0 & r_2 \bm{F}_2 \end{bmatrix}
\end{align}

Then we can rewrite the model as
\begin{align}
\bm{y} &= \bm{X}\bm{\delta} + \bm{W}_\ast \bm{\eta}_\ast + \bm{\epsilon} \\
p(\bm{\eta}_\ast | \tau_\epsilon, r_1, r_2) &= p(\bm{\eta}_1 | \tau_\epsilon, r_1)p(\bm{\eta}_2 | \tau_\epsilon, r_2) \\
&\propto \tau_\epsilon^{\rank(\bm{F}_\ast)/2}r_1^{\rank(\bm{F}_1)/2}r_2^{\rank(\bm{F}_2)/2}\exp\left\{-\frac{\tau_\epsilon}{2}\bm{\eta}_\ast^\top \bm{F}_\ast \bm{\eta}_\ast \right\}.
\end{align}
Note that $\rank(\bm{F}_\ast) = \rank(\bm{F}_1)+\rank(\bm{F}_2)$, and $\bm{W}_\ast$ has orthonormal columns because $\bm{W}_1$ and $\bm{W}_2$ have orthonormal columns and ${\cal C} (\bm{W}_1) \bot {\cal C} (\bm{W}_2)$.

We know that $\text{Var}(\bm{\delta}) = (\bm{X}^\top \bm{X})^{-1} \mathrm{E}[\sigma_\epsilon^2|\bm{y}]$ where $\sigma^2_\epsilon = 1/\tau_\epsilon$, so it suffices to show that $\mathrm{E}[\sigma_\epsilon^2|\bm{y}] \leq \mathrm{E}[\sigma^2_{\epsilon,NN}|\bm{y}]$. We find the posterior and conditional posterior distributions,
\begin{align}
    \sigma^2_{\epsilon,NN} | \bm{y} &\sim \text{inverse-gamma}\left(a_\epsilon + (n - p)/2, \frac{1}{b_\epsilon} + \frac{1}{2} \bm{y}^\top \bm{P}_{X^\bot} \bm{y} \right) \\
    \sigma^2_\epsilon | \bm{y}, r_1, r_2 &\sim  \text{inverse-gamma}\left(\begin{array}{l}a_\epsilon + a_1 + a_2 + (n - p)/2, \\ \frac{1}{b_\epsilon} + \frac{r_1}{b_1} + \frac{r_2}{b_2} + \frac{1}{2} \bm{y}^\top (\bm{P}_{X^\bot} - \bm{W}_\ast (\bm{I} + \bm{F}_\ast)^{-1} \bm{W}_\ast^\top) \bm{y} \end{array}\right)
\end{align}

Then,
$$\mathrm{E}\left[\sigma^2_\epsilon | \bm{y}, r_1, r_2\right] = \frac{\frac{1}{b_\epsilon} + \frac{r_1}{b_1} + \frac{r_2}{b_2} + \frac{1}{2} \bm{y}^\top (\bm{P}_{X^\bot} - \bm{W}_\ast (\bm{I} + \bm{F}_\ast)^{-1} \bm{W}_\ast^\top) \bm{y}}{a_\epsilon + a_1 + a_2 + (n - p)/2 - 1}.$$

Using total expectation,

$$\mathrm{E}\left[\sigma^2_\epsilon | \bm{y}\right] = \frac{\frac{1}{b_\epsilon} + \frac{\mathrm{E}[r_1|\bm{y}]}{b_1} + \frac{\mathrm{E}[r_2|\bm{y}]}{b_2} + \frac{1}{2} \bm{y}^\top \bm{P}_{X^\bot} \bm{y} - \mathrm{E}\left[\frac{1}{2} \bm{y}^\top \bm{W}_\ast (\bm{I} + \bm{F}_\ast)^{-1} \bm{W}_\ast^\top \bm{y}|\bm{y}\right]}{a_\epsilon + a_1 + a_2 + (n - p)/2 - 1}.$$

Now define
$$M(x) := \frac{\frac{1}{b_\epsilon} + \mathrm{E}[r_1|\bm{y}]x + \frac{b_1\mathrm{E}[r_2|\bm{y}]}{b_2}x + \frac{1}{2} \bm{y}^\top \bm{P}_{X^\bot} \bm{y}}{a_\epsilon + a_1b_1x + a_2b_1x + (n - p)/2 - 1},$$
and similary note that $M(0) = E[\sigma^2_{\epsilon, NN}|\bm{y}]$ and $M(1/b_1) \leq E[\sigma^2_{\epsilon}]$. We want to find conditions such that $M$ is decreasing in $x$ on $[0, 1/b_1]$.
\begin{align}
\frac{\partial M}{\partial x}
&= \frac{(a_\epsilon + a_1b_1x + a_2b_1x + (n - p)/2 - 1)(\mathrm{E}[r_1|\bm{y}] + \frac{b_1\mathrm{E}[r_2|\bm{y}]}{b_2})}{(a_\epsilon + a_1b_1x + a_2b_1x + (n - p)/2 - 1)^2} \nonumber \\
&\quad\quad - \frac{(\frac{1}{b_\epsilon} + \mathrm{E}[r_1|\bm{y}]x + \frac{b_1\mathrm{E}[r_2|\bm{y}]}{b_2}x + \frac{1}{2} \bm{y}^\top \bm{P}_{X^\bot} \bm{y})(a_1b_1 + a_2b_1)}{(a_\epsilon + a_1b_1x + a_2b_1x + (n - p)/2 - 1)^2}\\
&= \frac{(a_\epsilon + (n - p)/2 - 1)(\mathrm{E}[r_1|\bm{y}] + \frac{b_1\mathrm{E}[r_2|\bm{y}]}{b_2}) - (\frac{1}{b_\epsilon} + \frac{1}{2} \bm{y}^\top \bm{P}_{X^\bot} \bm{y})(a_1b_1 + a_2b_1)}{(a_\epsilon + a_1b_1x + a_2b_1x + (n - p)/2 - 1)^2} \\
&= \frac{\left\{(\mathrm{E}[r_1|\bm{y}] + \frac{b_1\mathrm{E}[r_2|\bm{y}]}{b_2}) - E[\sigma^2_{\epsilon, NN}|\bm{y}](a_1b_1 + a_2b_1)\right\}(a_\epsilon + (n - p)/2 - 1)}{(a_\epsilon + a_1b_1x + a_2b_1x + (n - p)/2 - 1)^2}.
\end{align}

So we have $\frac{\partial M}{\partial x} \leq 0$ if $(\mathrm{E}[r_1|\bm{y}] + \frac{b_1\mathrm{E}[r_2|\bm{y}]}{b_2}) / (a_1b_1 + a_2b_1) \leq E[\sigma^2_{\epsilon, NN}|\bm{y}]$, or equivalently if
$$\frac{\mathrm{E}[r_1|\bm{y}]/b_1 + \mathrm{E}[r_2|\bm{y}]/b_2}{\mathrm{E}[\tau_1]/b_1 + \mathrm{E}[\tau_2]/b_2} \leq E[\sigma^2_{\epsilon, NN}|\bm{y}]. $$

\qedsymbol

\subsection{Restricted Network Regression With a Single Random Effect}

The need for Theorem \plaintextref{2} is motivated by the fact that the model form in (\plaintextref{35})-(\plaintextref{37}) encompasses continuous restricted network models that include one additive sender or one additive receiver random effect, but not both. Taking the sender random effect as an example, we have obtain the restricted network regression model through the relations $\bm{\eta} = \bm{a}$, $\bm{W}=(\bm{I}-\bm{P}_X)\bm{A}$, $\bm{F}=\bm{I}_n$ (Table \ref{table:single-additive-effect-khan-mapping}). From the results in \citet{khan2020restricted}, any restricted regression model of the form in (\plaintextref{35})-(\plaintextref{37}) with a continuous response will also not alleviate spatial (or network) confounding according to Definition \plaintextref{1}.

\begin{table}
\caption{Table showing the mapping between the general form of spatial models studied in \citet{khan2020restricted}, (\plaintextref{35})-(\plaintextref{37}), and additive network models with a single sender random effect.}
\label{table:single-additive-effect-khan-mapping}
\centering
\begin{tabular}{ccl}
\toprule
Spatial & Restricted Network & Description \\
\midrule
$\bm{\eta}$  & $\bm{a}$ & Random effect\\
$\bm{W}$  & $(\bm{I} - \bm{P}_X)\bm{A}$ & Random effect design matrix\\
$\bm{F}$  & $\bm{I}$ & Random effect precision matrix \\
$\tau_s$  & $\sigma^{-2}_a$ & Random effect precision \\
\bottomrule
\end{tabular}
\end{table}

\section{Simulation of excess variation at specified canonical correlations} \label{apx:simulated-canonical-correlation}

The simulation studies in Section \plaintextref{4} depend on the ability to generate a vector of unobserved variation at a specified magnitude and a specified canonical correlation with a set of covariates. This is achieved through rescaling the parallel and orthogonal components of an initial vector through the following \texttt{R} function:

\begin{verbatim}
gencancor <- function(X, Y, rho) {
  PX <- X %*% MASS::ginv(t(X) %*% X) %*% t(X)
  Ypar <- c(PX %*% Y)
  Yperp <- Y - Ypar
  Ynew <- (rho * sqrt(sum(Yperp^2)) * Ypar +
      sqrt(1-rho^2) * sqrt(sum(Ypar^2)) * Yperp)
  return(Ynew * sqrt(sum(Y^2)) / sqrt(sum(Ynew^2)))
}
\end{verbatim}

This function works by taking an initial, i.i.d. normal random vector $Y$ and decomposing it into components parallel to and orthogonal to $X$. Those components are then scaled and recombined into a vector $Y'$, such that $Y'$ has the desired canonical correlation $\rho$ with $X$. The vector $Y'$ is then rescaled to have the same magnitude as the original $Y$ and returned.

\section{Eurovision Data} \label{sec:appendix-eurovision-data}

\begin{figure}
    \includegraphics[width=0.9\linewidth]{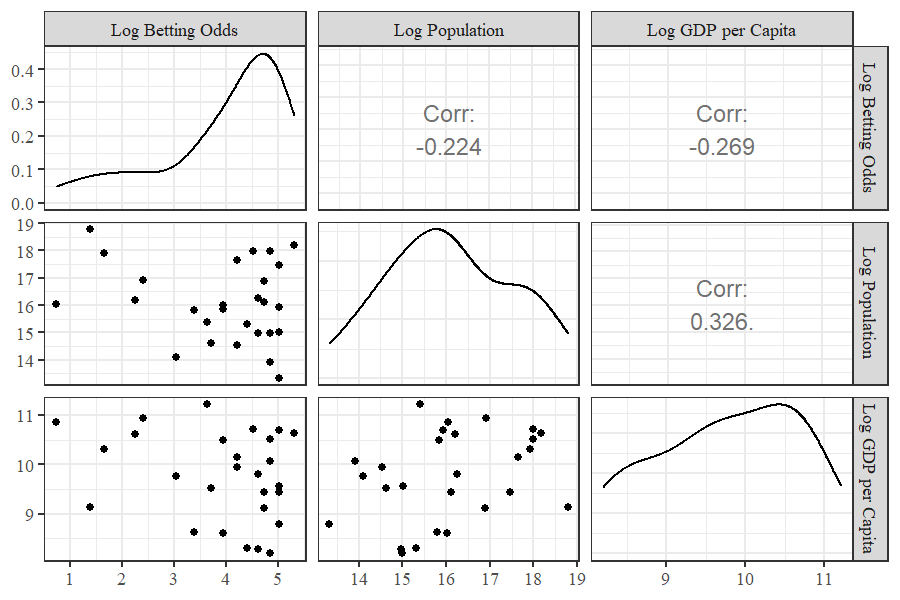}
    \caption{Pairs plot of the 3 receiver covariates in the Eurovision data analysis. There is low correlation between each pair of covariates. LogMedianOdds and LogGDP appear left skewed, LogPopulation is approximately normally-distributed.}
    \label{fig:eurovision-receiver-covariates}
\end{figure}

\begin{figure}
    \includegraphics[width=0.95\linewidth]{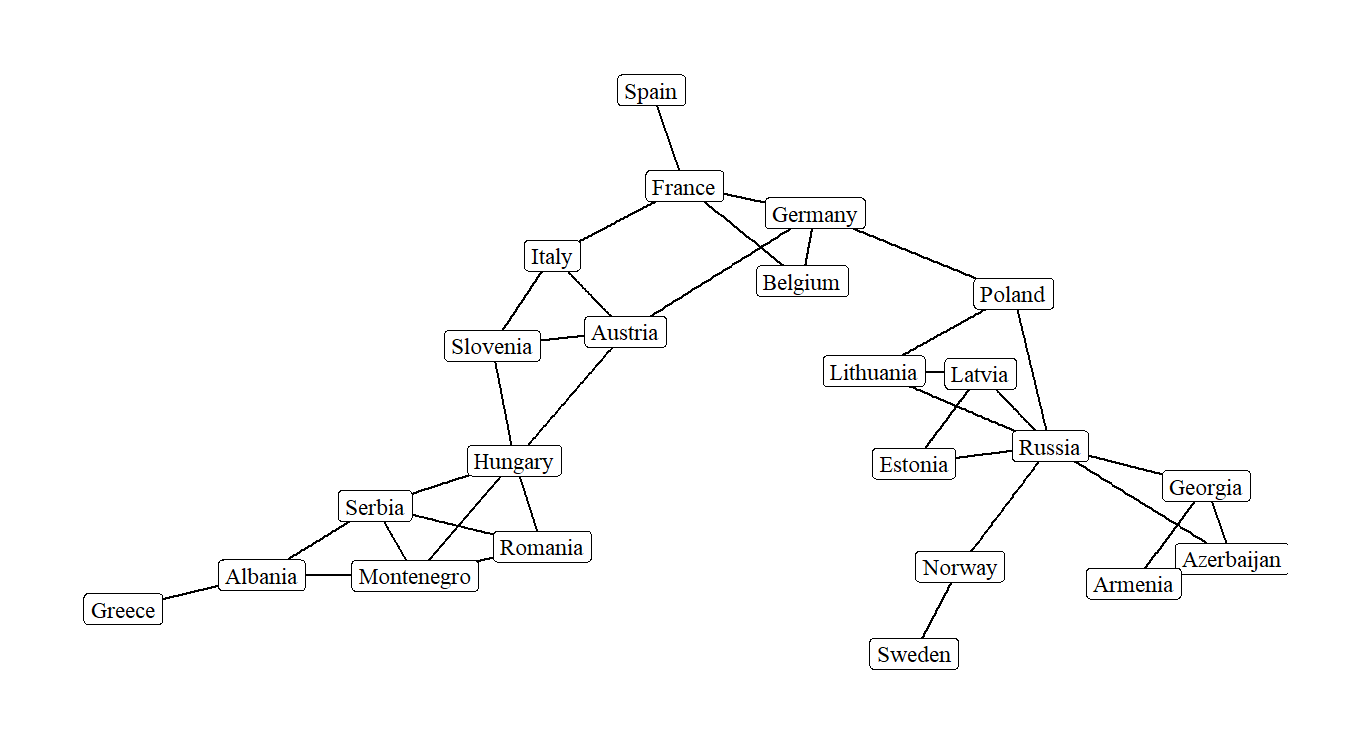}
    \caption{Eurovision country contiguity covariate illustrated as a network. Countries sharing a border are connected by an edge. Australia, Cyprus, Israel, and the United Kingdom border no other competing countries and are not shown.}
    \label{fig:eurovision-country-contiguity-network}
\end{figure}

Figures \ref{fig:eurovision-receiver-covariates} and \ref{fig:eurovision-country-contiguity-network} visualize the covariates in the Eurovision data analysis (Section \plaintextref{5}). Log-transforming the receiver covariates (Figure \ref{fig:eurovision-receiver-covariates}) results in no right skew. The receiver covariates are also not highly correlated with each other, avoiding multicollinearity in the model. As the Eurovision analysis found that country contiguity had a positive effect on voting, countries with large numbers of neighbors (e.g., Russia, Hungary) are more likely to receive votes, while countries with fewer neighbors (e.g., Australia, Cyprus, Israel, and the United Kingdom) are less likely to receive votes.

\bibliographystyle{apalike}
\bibliography{rnr.bib}